%% file: main.tex
\documentclass[letterpaper,twocolumn,10pt]{article}
\usepackage{usenix2019_v3}

\newcommand{\SYS}{\textsc{Paio}\xspace}
\newcommand{\SYSurl}{\url{https://github.com/dsrhaslab/paio}}

\usepackage{amssymb}
\usepackage{enumitem}
\usepackage{environ}
\usepackage{tabularx}
\usepackage{url}

\begin{document}

\date{}

\title{\Large \bf \SYS: A Software-Defined Storage Data Plane Framework}

\author{
{\rm Ricardo\ Macedo, Yusuke\ Tanimura$^{\dagger}$, Jason\ Haga$^{\dagger}$, Vijay\ Chidambaram$^{\ddagger}$, Jos\'{e}\ Pereira, Jo\~{a}o\ Paulo}\\
 INESC TEC \& University of Minho \\ 
 $^{\ \dagger}$National Institute of Advanced Industrial Science and Technology\\
 $^{\ \ddagger}$The University of Texas at Austin \& VMWare Research
} % end author

\maketitle

\input{input/00-abstract}
\input{input/01-introduction}

\input{input/02-challenges}
\input{input/03-design}
\input{input/04-implementation}
\input{input/05-controlapplications}
\input{input/06-evaluation}
\input{input/07-relatedwork}
\input{input/08-conclusion}

\input{input/09-acknowledgments}
\input{input/10-availability}

%-------------------------------------------------------------------------------
\bibliographystyle{plain}
\bibliography{bibliography}

\end{document}

%% file: input/00-abstract.tex
\begin{abstract}
We propose \SYS, the first general-purpose framework that enables system designers to build custom-made Software-Defined Storage (SDS) data plane stages. 
It provides the means to implement storage optimizations adaptable to different work\-flows and user-defined policies, and allows straightforward integration with existing applications and I/O layers.
\SYS allows stages to be integrated with modern SDS control planes to ensure holistic control and system-wide optimal performance.
We demonstrate the performance and applicability of \SYS with two use cases. 
The first improves $99^{th}$ percentile latency by 4$\times$ in industry-standard LSM-based key-value stores.
The second ensures dynamic per-application bandwidth guarantees under shared storage environments.  
\end{abstract}    

%% file: input/01-introduction.tex
\section{Introduction}
\label{sec:introduction}

Data-centric systems such as databases, key-value stores (KVS), and machine learning engines, share the need for efficient data storage and retrieval. 
This has led to the implementation of isolated I/O optimizations (\emph{e.g.,} scheduling, differentiation, caching) to address their storage requirements, such as resource fairness and throughput/latency SLOs~\cite{SILK:2019:Balmau,SplitIOScheduling:2015:Yang,Libra:2014:Shue}.
% \extrafootertext{\noindent\copyrightIEEE}
%
This approach, however, has two main drawbacks.
First, I/O optimizations are tightly integrated within the core of each solution, ma\-king it challenging to port these to other systems with similar performance goals.
Second, in shared environments where multiple systems operate concurrently and compete for shared resources, individual optimizations can conflict with each other~\cite{EnlighteningIOPath:2017:Kim}, leading to I/O contention and performance variation~\cite{Pisces:2012:Shue, PerformanceIsolation:2015:Xavier}.

The Software-Defined Storage (SDS)~\cite{IOFlow:2013:Thereska, SDSsurvey:2020:Macedo} paradigm promises an appealing solution to these limitations.
It aims at decoupling I/O functionality into two planes: \emph{control} and \emph{data}. 
The control plane is a logically centralized entity with system-wide visibility that enforces end-to-end policies in the I/O stack, which can be composed of different I/O layers (\emph{e.g.,} applications, databases, file systems, object stores) and physical storage devices (\emph{e.g.,} NVMe, SSD, HDD). 
Control algorithms, built on top of it, define the policies to be enforced at the I/O stack and generate rules directly applicable at the data plane. 
Examples of such control algorithms are used for achieving QoS provisioning~\cite{IOFlow:2013:Thereska, PriorityMeister:2014:Zhu}, performance control~\cite{sRoute:2016:Stefanovici, Moirai:2015:Stefanovici}, and resource fairness~\cite{Retro:2015:Mace, Pisces:2012:Shue}.

The data plane is a multi-stage component distributed over the I/O stack. 
Each data plane stage (or \emph{stage}, for short) implements custom I/O logic to apply over requests to meet a given policy. 
In particular, stages can provide simple data transformations such as encryption and compression~\cite{SafeFS:2017:Pontes, Crystal:2017:Gracia}, or more complex mechanisms such as token-buckets, I/O schedulers, and load balancers~\cite{IOFlow:2013:Thereska, Retro:2015:Mace, Pisces:2012:Shue, Malacology:2017:Sevilla}.

However, current SDS systems including IOFlow~\cite{IOFlow:2013:Thereska}, Retro~\cite{Retro:2015:Mace}, Crystal~\cite{Crystal:2017:Gracia}, and SafeFS~\cite{SafeFS:2017:Pontes}, are designed for enforcing policies over a \emph{specific set of layers} such as file systems, object stores, and hypervisors, or \emph{storage contexts} (\emph{e.g.,} cloud-based virtualization and application-specific storage stacks), thus limiting their adoption and applicability. 

In fact, introducing the ideas behind SDS over existing I/O layers, without significant system rewrite, is a challenging endeavor.
Layers interact with each other through rigid interfaces that cannot be extended with ease.
For example, the POSIX interface does not allow differentiating requests from different layers, or even \emph{workflows} of the same layer (\emph{e.g.,} background and foreground tasks of a KVS~\cite{SILK:2019:Balmau}).\footnote{We refer to the term \emph{``workflow''} as the connection between two I/O layers through where requests are transmitted.} 
Thus, intercepting and propagating request information to a stage is challenging, and its absence limits the context and granularity at which workflows can be differentiated and optimized.
Without this knowledge, optimizations must again be implemented individually at the layer, inhibiting code reuse and holistic tuning.

Solving these challenges requires a fundamental new abstraction, where the development of I/O optimizations should be made over a programmable and adaptable environment.
As such, we propose \SYS, the first general-purpose SDS data plane framework that enables system designers to build custom-made data plane stages.\footnote{\SYS\ stands for \textbf{P}rogrammable and \textbf{A}daptable \textbf{I}/\textbf{O}.}
By promoting code reuse and straightforward integration with I/O layers, \SYS eases the implementation of complex storage mechanisms that can adapt to different workflows and policies. 
The chief insight behind our work is that if we are able to intercept and differentiate requests as they flow through different layers, we can enforce policies without significantly changing the layers themselves.

\SYS makes this possible through three logical components.
First, a differentiation component classifies and differentiates requests at different levels of granularity. 
Leveraging from \emph{context propagation} ideas~\cite{ContextPropagation:2018:Mace}, \SYS propagates additional information of a given layer to the stage, enabling per-tenant and per-context (\emph{e.g.,} foreground and background tasks) differentiation.
Second, \SYS abstracts complex storage mechanisms into self-contained, custom-made \emph{enforcement objects}, which are programmable components that contain the I/O logic to apply over requests (\emph{e.g.,} token-buckets, I/O schedulers).
Third, \SYS exposes a control interface that allows SDS control planes to manage, monitor, and dynamically adapt each data plane stage.

We validate \SYS under two use cases.
% Use case 1
First, we implement a \SYS stage in RocksDB~\cite{RocksDB}, an industry-standard Log-Structured Merge tree (LSM) KVS, and demonstrate how to prevent latency spikes by orchestrating foreground and background tasks.
Results show that a \SYS-enabled RocksDB improves $99^{th}$ percentile latency by 4$\times$ under different workloads when compared to baseline RocksDB, and achieves similar tail latency performance when compared to SILK, a state-of-the-art, latency-oriented KVS~\cite{SILK:2019:Balmau}. 
% Use case 2
Second, we apply \SYS to TensorFlow~\cite{TensorFlow:2016:Abadi} and show how to achieve dynamic per-application bandwidth guarantees under a real shared storage scenario at the \emph{ABCI} supercomputer.\footnote{AI Bridging Cloud Infrastructure (\url{https://abci.ai}).}
Results show that all \SYS-enabled TensorFlow instances are provisioned with their bandwidth goals.

\SYS is implemented as a user-level library so de\-ve\-lo\-pers can create new data plane stage implementations and integrate them in different layers --- porting RocksDB and TensorFlow to a \SYS-enabled environment only required adding 85 and 22 lines of code, respectively.
Moreover, while this paper focuses on porting \emph{existing} I/O layers to an SDS-enabled environment, \SYS can also be used to simplify the development of \emph{future} data-centric systems.

% Contributions (itemized)
In sum, the paper makes the following contributions:

$\sbullet[.85]$ \SYS, a novel open-source data plane framework for building programmable and dynamically adaptable stages tailored for user-defined policies~(\cref{sec:design}-\cref{sec:impl}).
\SYS is available at \SYSurl.

$\sbullet[.85]$ The implementation of two data plane stages using \SYS, along with the corresponding control algorithms, to \emph{(1)} prevent latency spikes in industry-standard KVS, and \emph{(2)} to achieve per-application bandwidth guarantees under real shared storage deployments~(\cref{sec:controlapplications}).

$\sbullet[.85]$ Experimental results demonstrating the performance, applicability, and feasibility of \SYS under both synthetic and realistic scenarios~(\cref{sec:evaluation}).

%% file: input/02-challenges.tex
\section{Challenges}
\label{sec:challenges}

Modern infrastructures are made of multiple independent I/O layers that operate concurrently over the same resources.
To address the storage requirements specific to each layer, these implement system-specific and isolated I/O optimizations.
This design however, raises several challenges.

\begin{figure}[t]
    \centering
    \includegraphics[width=1\linewidth,keepaspectratio]{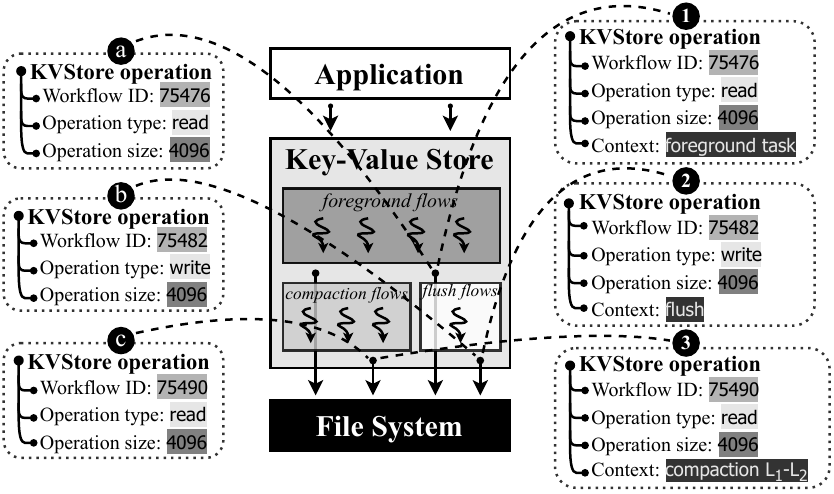}
    \vspace*{-15pt}
    % caption
    \caption{\textbf{Operations submitted from different I/O workflows.} 
    \emph{Example of the operation flow of a multi-layered I/O stack. 
    Left side depicts regular information extracted from operations between the KVS and File System, while the right side includes additional request information made available through context propagation.}}
    \label{fig:context-propagation}%
    \vspace*{-5pt}
\end{figure}

\paragraph{Tightly coupled optimizations}
I/O optimizations implemented at data-centric systems, such as caching, tiering, and scheduling, are single-purposed as they are tightly integrated within the core of each system. 
Implementing these optimizations requires deep understanding of the internal operation model and significant system rewrite, reducing their portability and adoption across systems that share similar principles.
For instance, porting SILK's I/O scheduler~\cite{SILK:2019:Balmau} to improve the tail latency performance of Le\-vel\-DB~\cite{leveldb:2020} and PebblesDB~\cite{PebblesDB:2017:Raju} is not trivial, and requires profound system refactoring.
As such, I/O optimizations should be disaggregated from the system's internal logic and moved to a dedicated layer, becoming generally applicable and portable across different scenarios.

\paragraph{Rigid interfaces}
The operation model of conventional I/O stacks requires layers to communicate through rigid interfaces that cannot be easily extended, 
discarding information that could be used to classify and differentiate requests at different levels of granularity.
For instance, consider the I/O stack depicted in Fig.~\ref{fig:context-propagation} made of an \emph{Application}, a \emph{KVS}, and a POSIX-compliant \emph{File System}.
POSIX operations submitted from the \emph{KVS} can be originated from different workflows, including foreground ({\circledparfootnotesize{a}}) and background flows \emph{i.e.,} flushes ({\circledparfootnotesize{b}}) and compactions ({\circledparfootnotesize{c}}).
From the \emph{File System}'s perspective however, it can only observe the size and type of a request, making it impossible to infer its origin. 
For example, {\circledparfootnotesize{a}} and {\circledparfootnotesize{c}} represent two 4~KiB-sized \texttt{read} operations that are originated from different contexts.
This loss of granularity reduces the possibility to differentiate and enforce complex policies over requests. 
Thus, layers should have access to additional request information to classify, differentiate, and enforce policies at a finer granularity.
Considering the previous example, by propagating the context that has originated a given request, we can pinpoint each operation to its origin and handle it accordingly.
Specifically, each request is now accompanied with a \emph{context} field that determines the origin of a request, namely a foreground task for {\circledparfootnotesize{1}} and compaction task for {\circledparfootnotesize{3}}.

\paragraph{Partial visibility}
I/O optimizations are implemented in isolation and are oblivious of the remaining layers of the I/O stack.
Under this design, layers compete for shared resources, leading to conflicting optimizations, misconfigurations, I/O contention, and performance variation.
As such, optimizations should have system-wide visibility to ensure coordinated and holistic control of all storage resources.

%% file: input/03-design.tex
\section{\SYS Design}
\label{sec:design}

\SYS is a general-purpose SDS framework that enables system designers to build custom-made data plane stages. 
A data plane stage built with \SYS allows the classification and differentiation of I/O requests, and the enforcement of different mechanisms according to user-defined storage policies.
Examples of such policies can be as simple as adjusting the workflows' rate of greedy tenants to achieve resource fairness, or more complex ones as coordinating the rate of foreground and background workflows to ensure sustained tail latency.
To achieve this, and to address the challenges pointed in \cref{sec:challenges}, \SYS's design is built over three core principles. 

\paragraph{Programmable and extensible building blocks}
The data plane must be programmable to allow developing stages tailored for each layer.
It should be extensible and provide the necessary abstractions for building custom I/O me\-cha\-nisms, such as caches, schedulers, and token-buckets, to employ over requests.
These properties are key for supporting a wide range of I/O mechanisms tailored for the requirements of different layers.

\paragraph{Fine-grained control over I/O} 
The data plane must have granular control over I/O workflows to classify and dif\-fe\-ren\-ti\-ate requests at different levels, such as per-application, per-workflow, or per-request type. 
This allows im\-ple\-men\-ting a rich set of policies over the I/O stack (\emph{e.g.,} QoS provisioning, resource fairness, load balancing).

\paragraph{Control interface}
The data plane should expose a control interface that abstracts the complexity of its internal organization, and allow the SDS control plane to dynamically adapt each stage to new policies and workload variations.

\begin{figure}[t]
    \centering
    \includegraphics[width=1\linewidth,keepaspectratio]{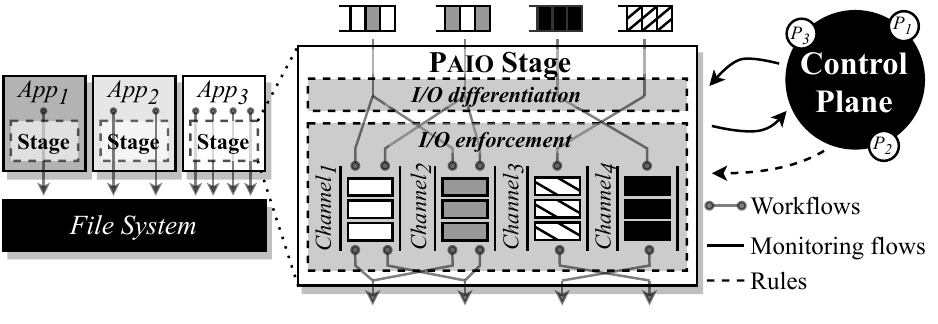}
    \vspace*{-15pt}
    \caption{\textbf{\SYS overview.} \SYS \emph{is a general-purpose SDS framework that allows implementing fine-tuned data plane stages at different points of the I/O stack.}}
    \label{fig:paio-overview}%
\end{figure}

\subsection{Abstractions in \SYS}
\label{subsec:abstractions}
\SYS uses four main abstractions, namely \emph{enforcement objects}, \emph{channels}, \emph{context}, and \emph{rules}.

\paragraph{Enforcement object}
An enforcement object is a self-con\-tai\-ned, single-purposed mechanism that contains custom I/O logic to apply over requests.
Examples of such mechanisms can range from \emph{performance control} and \emph{resource management} such as token-buckets, I/O schedulers, and caches, \emph{data transformations} as compression and encryption, to \emph{data management} (\emph{e.g.,} data prefetching, tiering).
This abstraction provides to system designers the necessary flexibility and extensibility for developing new I/O mechanisms tailored for enforcing specific storage policies over requests.

\paragraph{Channel}
A channel provides a stream-like abstraction through which requests flow.
Each channel contains one or more enforcement objects, as well as a \emph{rule} that maps requests to the respective enforcement object to be enforced.
The combination of channels and enforcement objects is designed to ease the implementation of new storage services, while promoting their reutilization and applicability. % to different applications and I/O layers.

\paragraph{Context}
A context represents a metadata-like object that contains a set of elements that characterize a request.
These elements (or \emph{classifiers}) include the \emph{workflow id} (\emph{e.g.,} thread-ID), \emph{request type} (\emph{e.g.,} \texttt{read}, \texttt{open}, \texttt{put}, \texttt{get)}, \emph{request size}, and the \emph{request context}, which defines the context of a request (\emph{e.g.,} foreground or background tasks, flush or compaction).
For each request, \SYS generates the corresponding \emph{context} object that is used for classifying, differentiating, and en\-for\-cing the request over the respective mechanisms. 

\paragraph{Rule}
In \SYS, a rule represents an action that updates the state of a data plane stage.
Rules are submitted by the control plane, and are organized in three types: \emph{housekeeping rules} manage the internal stage organization, \emph{differentiation rules} classify and differentiate I/O requests, \emph{enforcement rules} adjust enforcement objects upon workload variations.

\subsection{Architecture}
\label{subsec:architecture}

Fig.~\ref{fig:paio-overview} outlines \SYS's high-level architecture, which consists of data plane stages and an external control plane.
\SYS's design targets the workflows of any given point of the I/O stack. 
To orchestrate these, stages are embedded within layers to intercept requests and enforce user-defined policies.

To achieve this, \SYS is organized in four main components.
First, \SYS exposes an \emph{Instance interface} (\cref{subsec:interfaces}) that bridges the targeted layer (\emph{App$_3$}) and the data plane stage. 
It intercepts all requests that are destined to the next layer (\emph{App$_3$$\rightarrow$File System}) and generates a per-request \emph{context} object that contains all request's classifiers. 
Both request and \emph{context} object are then submitted to the data plane stage.

Second, a \emph{differentiation module} (\cref{subsec:differentiation}) classifies and dif\-fe\-ren\-ti\-ates requests based on their \emph{context} object.
Requests can be differentiated at different levels of granularity, being then dispatched to the correct channel to be enforced.

Third, \SYS provides an \emph{enforcement module} (\cref{subsec:enforcement}) that is responsible for enforcing policies over requests and is organized with several channels and enforcement objects.
For each request, the channel selects the enforcement object to employ its I/O mechanism.   
After being enforced, requests are returned to the original data path and submitted to the next I/O layer (\emph{File System}).

Finally, \SYS exposes a \emph{control interface} (\cref{subsec:interfaces}) that allows the SDS control plane to orchestrate the stage lifecycle, such as creating channels and enforcement objects, propagating new enforcement rules, and collecting I/O statistics.
Exposing such a control interface allows \SYS stages to be managed by existing SDS control planes~\cite{IOFlow:2013:Thereska,Crystal:2017:Gracia,Retro:2015:Mace}.

An external control plane orchestrates each stage to cope with user-defined policies.
It communicates with the data plane through \SYS's control interface to: submit \emph{rules}, either for internal management, differentiation, or fine-tuning enforcement objects; and monitoring, to keep track with the stage's performance and ensure that all policies are met.

\subsection{I/O Differentiation}
\label{subsec:differentiation}  

\input{input/floaters/table-differentiation-examples}

\SYS's differentiation module provides the means to classify and differentiate requests at different levels of granularity, namely \emph{per-workflow}, \emph{request type}, and \emph{request context}.
The process for differentiating requests is done in two phases.

The first phase, which happens at startup time, defines how requests should be differentiated and which requests a channel receives.
To do so, first \SYS specifies the \emph{context}'s classifiers that will be considered at runtime, which can be a single classifier or a combination of them.
For example, to use per-workflow differentiation, \SYS only considers the \emph{context's workflow id}, while to differentiate requests based on their context and type, \SYS considers both \emph{request context} and \emph{request type} classifiers.
Second, \SYS attributes specific \emph{context} classifiers to each channel, which are used to map requests to the respective channel.
Namely, it defines the exact request's \emph{workflow id}, \emph{context}, and/or \emph{type} that a channel receives.
Table~\ref{table:diff} provides examples of this attribution of classifiers: \emph{channel$_1$} only receives requests from \emph{flow$_1$}, while \emph{channel$_2$} only handles \texttt{read} requests originated from background tasks. \emph{Channel$_3$} receives compaction-based \texttt{write} requests from \emph{flow$_5$}. 
To generate a unique identifier that maps requests to channels, \emph{context}'s classifiers can be concatenated into a single string or hashed into a fixed-size token (\cref{subsec:implementation}).
These configurations can be set up either through \emph{differentiation rules} submitted by the control plane or directly configured at the data plane stage. 

The second phase, which happens at execution time, differentiates and forwards requests to the respective channel.
This is achieved in three steps: \emph{context propagation}, \emph{channel selection}, and \emph{enforcement object selection}.

\begin{figure}[t]
    \centering
    \includegraphics[width=1\linewidth,keepaspectratio]{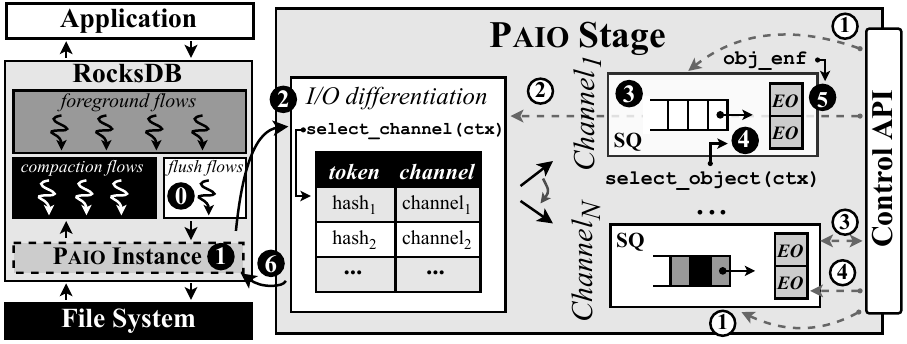}
    \vspace*{-15pt}
    % Caption
    \caption{\textbf{\SYS operation flow.}
    \emph{Black filled circles depict the execution flow of a request in the \SYS stage (\circledpar{0} -- \circledpar{6}). White filled circles depict the flow of control operations submitted from the SDS control plane to the data plane stage (\emptycircledpar{1} -- \emptycircledpar{4}).}
    }
    \label{fig:paio-operation-flow}%
    \vspace*{-5pt}
\end{figure}

\paragraph{Context propagation}
Several request's classifiers, such as \emph{workflow id}, \emph{type}, and \emph{size}, can be extracted with ease from observing raw I/O requests.
However, additional request information that is only known to the layer that is ma\-na\-ging its lifecycle could be used to expand the policies to be enforced over the I/O stack (\cref{sec:challenges}). 
An example of such information, as depicted in Fig.\ref{fig:context-propagation} and Table~\ref{table:diff}, is the \emph{operation context}, which allows understanding the original context of a given request, specifically if it is a foreground or background task, a flush or compaction operation, or other.

Thus, \SYS enables the propagation of additional request information from the targeted system to the data plane stage.
\SYS borrows ideas from \emph{context propagation} -- a commonly used technique in distributed systems monitoring that enables a system to forward context along the execution path~\cite{ContextPropagation:2018:Mace,PivotTracing:2018:Mace} -- and applies them to enhance I/O differentiation and enforcement.
To achieve this, systems designers need to instrument the critical data path of the targeted layer where the information can be collected.
The information is then propagated to the \emph{Instance interface} and included as the \emph{request's context} classifier on the creation of the \emph{context} object.

For example, consider the I/O stack in Fig.~\ref{fig:paio-operation-flow} made of an \emph{Application}, \emph{RocksDB} KVS, a \SYS stage, and a POSIX-compliant \emph{File System}. 
\emph{RocksDB}'s background workflows are responsible for serving flush and compaction jobs.
Before being submitted to the \emph{File System}, jobs are translated into several POSIX \texttt{read} and \texttt{write} operations, leading to a loss of granularity at the operation context.
To propagate this information, system designers instrument \emph{RocksDB}'s critical path responsible for managing flush or compaction jobs {\circled{0}} to capture their operation context (expressed as \texttt{bg\_flush} or \texttt{bg\_compaction}).
This information is then propagated to the \SYS \emph{Instance}, where the \emph{context} object will be created with all classifiers {\circled{1}} and submitted to the \SYS stage.

Note that this process is not mandatory, as it can be skipped for stages whose policies can be met without the propagation of additional request information (\cref{subsec:qos-diff}).

\paragraph{Channel selection}
As depicted in Fig.~\ref{fig:paio-operation-flow} {\circled{2}}, for each in\-co\-ming request, \SYS selects the channel that must service it.
To do so, \SYS invokes a \texttt{select\_channel} call that verifies the \emph{context}'s classifiers and maps the request to the respective channel to be enforced.
This mapping is done as described in the first phase of the differentiation process.

\paragraph{Enforcement object selection}
As each channel may contain several enforcement objects, analogously to channel selection, \SYS selects the correct object to service the request.  
As depicted in Fig.~\ref{fig:paio-operation-flow} {\circled{4}}, for each request, the channel invokes a \texttt{select\_object} call that verifies the \emph{context} classifiers and maps the request to the respective enforcement object, which will then employ its I/O mechanism (\cref{subsec:enforcement}).

\subsection{I/O Enforcement}
\label{subsec:enforcement}

The enforcement module provides the building blocks for developing fine-tuned I/O services to be employed over workflow requests.
It is composed of several channels, each containing one or more enforcement objects.
The enforcement process begins after the channel selection.
As depicted in Fig.~\ref{fig:paio-operation-flow}, requests are forwarded to the selected channel and placed in a \emph{submission queue} {\circled{3}}.
For each request, \SYS invokes the \texttt{select\_object} call that selects the enforcement object to use {\circled{4}} and applies its I/O mechanism over the request {\circled{5}}. 
Examples of such mechanisms include token-buckets, caches, prefetching, and encryption schemes (we discuss how to build enforcement objects in the following paragraphs).
Since several mechanisms can change the request's original state, such as \emph{data transformations} (\emph{e.g.,} encryption, compression), during this phase, the enforcement object generates a \emph{result} object to store the updated request. 
This \emph{result} is then returned to the \SYS \emph{Instance}, that will unmarshall and forward it to the original data path {\circled{6}}.

Depending on the policies and I/O mechanisms to be employed, \SYS can enforce requests by only using \emph{context} objects.
While data transformations are directly applicable over the request's content, performance control mechanisms such as token-buckets and schedulers, only require specific request's metadata to be enforced (\emph{e.g.,} request type, size, priority).
Thus, to avoid adding unnecessary overhead to the system execution, upon the submission of requests to the stage, \SYS allows for the  request's content to be copied to the stage's execution path only when necessary (\emph{e.g.,} enforcing data transformation mechanisms).

A key feature of \SYS is that the targeted system is oblivious to the enforcement of its requests, as well as the number of channels and I/O mechanisms in the stage. 

\input{input/floaters/table-interface-definitions}

\paragraph{Building enforcement objects}
As depicted in Table~\ref{table:interfaces}, \SYS exposes to system de\-sig\-ners a simple API that allows building enforcement objects. 
% initialize
An \texttt{obj\_init(s)} call creates and configures an enforcement object with initial state \texttt{s}.
% configure
\texttt{obj\_config(s)} provides the tuning knobs to update the enforcement object's internals with a new state \texttt{s}. 
It enables the control plane to dynamically adapt the enforcement object to workload variations and new policies.
% enforce
An \texttt{obj\_enf(ctx,r)} call, or \emph{``object enforce''}, implements the actual I/O logic to apply over requests.
It returns a \emph{result} object that contains the updated request (\texttt{r}) after applying its logic.
It also receives a \emph{context} object (\texttt{ctx}) that is used to employ different actions over the I/O request.

To demonstrate the use of this abstraction, we focus on the implementation of a token-bucket mechanism~\cite{NetworkCalculus:2001:Boudec}. 
We use the token-bucket to control the rate and burstiness of I/O workflows. 
Each workflow is served at a given \emph{token rate}.
The bucket is configured with a \emph{bucket size}, which delimits the maximum token capacity, and a \emph{refill period} that defines the period to replenish the bucket.
On \texttt{obj\_init} the bucket is created and its size and refill period are set.
Upon each request, the \texttt{obj\_enf} call verifies the \emph{context's size} classifier and computes the number of tokens to be consumed, so the request can proceed. 
If not enough tokens are available, the request waits for the bucket to be refilled.
Upon workload variations, the control plane may need to adjust the token rate, triggering a \texttt{obj\_config} that adjusts the bucket size and/or refill period.

By default, \SYS preserves the operation logic of the targeted system (\emph{e.g.,} ordering, error handling), as both enforcement objects and operations submitted to \SYS follow a synchronous model.
While developing asynchronous enforcement objects is feasible, one needs to ensure that both correctness and fault tolerance guarantees are preserved.

%% file: input/floaters/table-differentiation-examples.tex
\begin{table}[t]
    \centering
    % Caption
    \caption{\emph{Examples of the type of requests a channel receives.}}
    \small
    \vspace*{-5pt}
    \begin{tabular}{lccc}
        \toprule
        \textbf{Channel}        & \textbf{Workflow ID}  & \textbf{Request context}  & \textbf{Request type}  \\ \midrule
        \emph{channel$_1$}      & \emph{flow$_1$}       & ---                       & ---   \\
        \emph{channel$_2$}      & ---                   & \emph{background tasks}   & \texttt{read}  \\
        \emph{channel$_3$}      & \emph{flow$_5$}       & \emph{compaction}         & \texttt{write} \\ \bottomrule
    \end{tabular}
    \label{table:diff}
    \vspace*{-5pt}
\end{table}

%% file: input/floaters/table-interface-definitions.tex
\begin{table}[t]
    \centering
    \caption{\emph{Interface definitions of \SYS modules.}}
    \vspace*{-5pt}
    \footnotesize
    \begin{tabular}{p{3pt} l l}
    \toprule
    \multirow{5}{*}{\textbf{1}$^*$ }    & \texttt{\textbf{stage\_info()}}   & Get data plane stage information                              \\
                                        & \texttt{\textbf{hsk\_rule(t)}}    & Housekeeping rule with tuple \texttt{t}                       \\
                                        & \texttt{\textbf{dif\_rule(t)}}    & Differentiation rule with tuple \texttt{t}                    \\ 
                                        & \texttt{\textbf{enf\_rule(id,s)}} & Enf. rule over enf. object \texttt{id} with state \texttt{s}  \\
                                        & \texttt{\textbf{collect()}}       & Collect statistics from data plane stage                      \\ \midrule
    {\textbf{2}$^\dagger$}              & \texttt{\textbf{enforce(ctx,r)}}  & Enforce context \texttt{ctx} and request \texttt{r}             \\ \midrule
    \multirow{3}{*}{\textbf{3}$^\star$} & \texttt{\textbf{obj\_init(s)}}    & Initialize enf. object with state \texttt{s}                  \\
                                        & \texttt{\textbf{obj\_enf(ctx,r)}} & Enforce I/O mechanism over \texttt{ctx} and \texttt{r}          \\
                                        & \texttt{\textbf{obj\_config(s)}}  & Configure enf. object with state \texttt{s}                   \\ \bottomrule
    \multicolumn{3}{l}{$^*$\textbf{Control API}; $^\dagger$\textbf{Instance API}; $^\star$\textbf{Enforcement object API}.}
\end{tabular}
\label{table:interfaces}
% \vspace*{-5pt}
\end{table}

%% file: input/04-implementation.tex
\section{\SYS Interfaces and Implementation}
\label{sec:impl}

We now detail how \SYS interacts with control planes and I/O layers, what a typical operation flow looks like in a \SYS stage, and how \SYS's prototype is implemented.

\subsection{Interfaces}
\label{subsec:interfaces}

\paragraph{Control interface}
Communication between \SYS stages and the control plane is achieved by exposing five API calls, depicted in Table~\ref{table:interfaces}.
A \texttt{stage\_info} call lists information about the stage, including the \emph{process identifier} (PID), \emph{stage identifier}, and the number of intercepted workflows.

Rule-based calls are designed to directly orchestrate \SYS's internal mechanisms.
For data plane maintenance, it defines \emph{housekeeping rules} (\texttt{hsk\_rule}) that manage the stage lifecycle (\emph{e.g.,} create channels and enforcement objects), and \emph{differentiation rules} (\texttt{dif\_rule}) that map requests to channels and enforcement objects. 
\emph{Enforcement rules} (\texttt{enf\_rule}) dynamically adjust the internal state (\texttt{s}) of a given enforcement object (\texttt{id}) upon workload and policy variations. 

To ensure policies are met at any given time, the control plane continuously monitors stages. 
A \texttt{collect} call gathers key performance metrics of all workflows (\emph{e.g.,} IOPS, bandwidth) to adjust enforcement objects to workload variations.

Through this interface, the control plane is able to define how \SYS stages handle I/O requests.
Nonetheless, concerns related to the coordination and dependability of data plane stages, as well as the resolution of conflicting policies and layers are responsibility of the control plane~\cite{SDSsurvey:2020:Macedo}, and are thus orthogonal to this paper.

\paragraph{Instance interface}
Communication between a layer and a stage is made through the \emph{Instance} interface, that establishes the connection between workflows and \SYS's internal mechanisms.
As depicted in Table~\ref{table:interfaces}, it provides an \texttt{en\-for\-ce} call that intercepts requests from the layer and forwards them, along the associated \emph{context} object, to the stage.
To select where requests should be intercepted, system de\-sig\-ners need to instrument the layer's critical path that invokes the next layer's calls.
For example, to orchestrate POSIX \texttt{read} operations of a given application, they need to be intercepted before being submitted to the file system. 
Here, the \emph{context} object is generated and submitted alongside the request to the \SYS stage through \texttt{enforce}.
After enforcing the request, the original data path execution is resumed.

To simplify layer instrumentation, \SYS also provides layer-oriented interfaces (\emph{e.g.,} POSIX, key-value), so users only need to replace the original call for a \SYS one.

\subsection{A Day in the Life of a Request}
\label{subsec:operationflow}

To illustrate how workflows are orchestrated by \SYS, we consider the I/O stack depicted on Fig.~\ref{fig:paio-operation-flow}, and consider the enforcement of the following policy: \emph{``limit the rate of RocksDB flush operations to X MiB/s''}. 
RocksDB generates foreground and background flows, containing client-based operations and internal maintenance work (\emph{e.g.,} flushes, compactions), respectively.
Before execution, the system designer instruments RocksDB for context propagation {\circled{0}} and redirecting flush-based requests to \SYS\xspace {\circled{1}}.
In \circledpar{0}, the path responsible for managing flush jobs is instrumented to capture the operation context, expressed as \texttt{bg\_flush}.
In \circledpar{1} RocksDB flushes are translated into several POSIX \texttt{write} requests, and the \SYS \emph{Instance} only handles these. 

At startup time, RocksDB initializes the \SYS stage, which connects to an already deployed control plane, and identifies itself with \texttt{stage\_info}.
The control plane then submits \texttt{hsk\_rule}s \emptycircled{1} to create a channel and an enforcement object that contains a token-bucket mechanism whose rate is set to $X$ MiB/s. 
Finally, it creates two \texttt{dif\_rule}s for the channel and enforcement object selection {\emptycircled{2}}.

At execution time, upon a flush-based \texttt{write} request, a \emph{context} object is created with its \emph{request type} (\texttt{write}), \emph{context} (\texttt{bg\_flush}), \emph{size}, and \emph{workflow id} (thread-ID), and submitted, along the request, to the stage through \texttt{enforce} {\circled{1}}.
Then, the stage selects the channel {\circled{2}} to be used and enqueues the request {\circled{3}}. 
The channel will then select the enforcement object to service the request {\circled{4}}.
On \texttt{obj\_enf} {\circled{5}}, the token-bucket consumes tokens from the bucket and generates the \emph{result}. 
If not enough tokens are available, the request waits until the bucket is refilled. 
After enforcing the request, the \emph{result} is returned to the \SYS \emph{Instance} {\circled{6}}, and the original \texttt{write} proceeds to the file system. 

During this time, the control plane continuously mo\-ni\-tors the data plane stage. 
On \texttt{collect}, the stage gathers performance metrics (\emph{e.g.,} throughput) and sends them to the control plane {\emptycircled{3}}.
Based on these statistics, the control plane adjusts the bucket's rate to ensure RocksDB's flush operations flow at $X$ MiB/s, generating \texttt{enf\_rule}s with new rates for the enforcement object to be adjusted {\emptycircled{4}}.

\subsection{Implementation}
\label{subsec:implementation}

We have implemented \SYS prototype with 8,000 lines of C++ code.
To enforce the policies targeted in our use cases, we implemented two enforcement objects -- \texttt{Noop} implements a pass-through mechanism (\emph{i.e.,} submits requests to the next layer without additional data processing), and \texttt{DRL} implements a token-bucket, whose goal is to \textbf{d}ynamically \textbf{r}ate \textbf{l}imit I/O requests. 
The rate at which the bucket serves requests is given in \emph{tokens/s}.
A \texttt{rate(r)} routine, used on \texttt{obj\_config}, changes the bucket's \emph{size} according to a function between \texttt{r} and \emph{refill period}.
On \texttt{obj\_enf} the bucket consumes \texttt{N} tokens. 
We consider that the cost of requests is constant \emph{i.e.,} each byte of a \texttt{read} or \texttt{write} request re\-pre\-sents a token. 
Although the I/O cost depends on several factors (\emph{e.g.,} workload, operation type, cache hits), we continuously calibrate the token-bucket so its rate converges to the policies' goal. 
Our experiments show that this approach works well in our scenarios, as the bucket's rate converges within few interactions with the control plane.
Nevertheless, determining the I/O cost is complementary to our work~\cite{Libra:2014:Shue, mClock:2010:Gulati}.
Combining \SYS with these could be useful under scenarios where policies are sensitive to the I/O cost.

\SYS implements per-workflow statistic counters at channels to register bandwidth of intercepted requests, number of operations, and mean throughput between collection periods.
On \texttt{collect}, it aggregates the statistics and reports them to the control plane.
Communication between the control plane and stages is established through UNIX Domain Sockets.
To create the unique identifiers that map requests to channels and enforcement objects, we used a computationally cheap hashing scheme~\cite{MurmurHash:2020:Appleby} (\emph{i.e.,} MurmurHash3) that hashes classifiers into a fixed-size token.

\SYS is provided to the community as an open-source user-space library, so developers can create new stage implementations and integrate them in different I/O layers, as shown in \cref{sec:controlapplications}.
Moreover, \SYS's design is aligned with the current efforts of moving the storage stack to user-level through kernel-bypass technologies (\emph{e.g.,} SPDK, PMDK).

\paragraph{Control plane}
We built a simple but fully-functional control plane with 3,600 lines of C++ code that enforces policies for the two use cases of this paper.
Policies were implemented as control algorithms.
To calibrate enforcement objects, besides stage statistics, it collects the I/O metrics generated at the targeted layer from the \texttt{/proc} file system~\cite{Proc:2020}.
Specifically, it inspects the \texttt{read\_bytes} and \texttt{write\_bytes} counters, which represent the number of bytes read/written from/to the block layer, and compares them with stage statistics to converge to the targeted performance goal.

%% file: input/05-controlapplications.tex
\section{Use Cases and Control Algorithms}
\label{sec:controlapplications}

We now present two use cases that showcase the applicability of \SYS for different applications and performance goals.

\subsection{Tail Latency Control in Key-Value Stores}
\label{subsec:tail-latency}

LSM KVSs such as LevelDB~\cite{leveldb:2020} and RocksDB~\cite{RocksDB} use \emph{foreground flows} to attend client requests, which are enqueued and served in FIFO order. 
\emph{Background flows} serve internal operations, namely flushes and compactions. 
Flushes are sequentially written to the first level of the tree ($L_0$) by a single thread and only proceed when there is enough space.
Compactions are held in an internal FIFO queue waiting to be executed by a dedicated thread pool.
Except for low level compactions ($L_0$--$L_1$), these can be executed in parallel.

\paragraph{Latency spikes}
A common problem of LSM KVSs is the interference between these workflows, generating latency spikes for client requests.
Latency spikes occur when flushes cannot proceed. 
First when $L_0$--$L_1$ compactions are slow, either because there is not enough disk bandwidth or because they are waiting in the compaction queue.
This increases the size of $L_0$, blocking flushes when there is no more storage quota left at this level. 
Second when flushes are slow, because there is not enough disk bandwidth for the operation to be executed timely.
These lead the memtable to fill, stalling client writes and causing latency spikes. 

SILK, a KVS built over RocksDB, addresses this with an I/O scheduler that: allocates bandwidth for internal operations when client load is low; prioritizes flushes and low level compactions, as they impact client latency; and preempts high level compactions with low level ones~\cite{SILK:2019:Balmau}. 
It employs these techniques with the following control algorithm.
As these KVSs are embedded, the KVS I/O bandwidth is bounded to a given rate (\emph{KVS}$_B$).
It monitors clients' bandwidth ($Fg$), and allocates any leftover bandwidth (\emph{left}$_B$) to internal operations ($I_B$), given by $I_B=$ \emph{KVS}$_B -$\emph{Fg}.  
To enforce rate $I_B$, SILK uses RocksDB's rate limiters~\cite{RateLimiterRocksDB:2020}.
Flushes and $L_0$--$L_1$ compactions have high priority and are provisioned with minimum I/O bandwidth ($min_B$).
High level compactions have low priority and can be paused at any time.
Because all compactions share the same thread pool, it is possible that, at some point, all threads are handling high level compactions. 
As such, SILK preempts one of them to execute low level compactions. 

However, implementing SILK's I/O optimizations over RocksDB required reorganizing its internal operation flow, changing core modules made of thousands of LoC including background operation handlers, internal queuing logic, and the thread pools allocated for internal work. 
Further, porting these optimizations to other KVSs that would equally benefit from them would require a substantial re-implementation effort. 
As such, we propose an alternative approach.

\paragraph{\SYS} 
Rather than modifying the RocksDB engine, we noticed that several of these optimizations could be achieved by orchestrating I/O workflows.
Thus, we implemented SILK's design principles in SDS fashion: a \SYS data plane stage implements the I/O mechanisms for prioritizing and rate limiting background flows, while the control plane re-implements SILK's I/O scheduling algorithm to orchestrate the stage, increasing the portability and applicability of these techniques over systems that share a similar design.

The stage intercepts all RocksDB workflows. 
We consider each RocksDB thread that interacts with the file system as a workflow.
Differentiation is made using the \emph{workflow id} and \emph{request context} classifiers.
We instrumented RocksDB to perform context propagation, which only required adding 47 LoC.
When a flush or compaction operation is triggered, the \emph{context} object is created with the respective \emph{request context} (\emph{e.g.,} \texttt{bg\_flush}, \texttt{bg\_compaction\_L$_{0}$\_L$_{1}$}).
Foreground flows are enforced with a \texttt{Noop} object that collects statistics of clients' bandwidth.
Background flows are forwarded to channels made of \texttt{DRL} enforcement objects. 
Flushes flow through a single channel. 
As compactions with different priorities can flow through the same channel, it contains two \texttt{DRL} objects configured at different rates, one for low level compactions and another for high level ones.
\SYS also collects the I/O bandwidth of flushes ($Fl$), and low level ($L_0$) and high level compactions ($L_N$).
Integrating \SYS in RocksDB required adding 85 LoC, as listed in Table~\ref{table:lines-of-code}.

\input{input/floaters/algorithm-kvs}

On the control plane we implemented a SDS version of the SILK's scheduling algorithm, as shown in Algorithm~\ref{alg:silk}.
The algorithm uses a feedback control loop that performs the following steps.
First, it collects statistics from the stage (\texttt{1}) and computes leftover disk bandwidth (\emph{left}$_B$) to assign to internal operations (\texttt{2}).
To ensure that background flows keep flowing, it defines a minimum bandwidth threshold (\texttt{3}), and distributes \emph{left}$_B$ according to workflows priorities (\texttt{4-11}).
It verifies if high priority tasks are executing, equally distributing \emph{left}$_B$ and assigning minimum bandwidth ($min_B$) to high level compactions (\texttt{5}). 
It is important that high level compactions keep flowing to prevent low level ones from being blocked in the queue.
If a single high priority task is being executed, \emph{left}$_B$ is allocated to it and $min_B$ to others (lines \texttt{6-9}).
It allocates \emph{left}$_B$ to low priority tasks, when high levels ones are not executing (\texttt{11}).
Then, it generates and submits \texttt{enf\_rule}s for adjusting the rate of each enforcement object (\texttt{12}).
It first assigns rate $B_{Fl}$ to those responsible for flushes.
For low priority compactions it splits $B_{L_N}$ between all \texttt{DRL} objects that handle these.
Because high priority compactions are executed sequentially, it assigns $B_{L_0}$ rate to the respective \texttt{DRL} objects.

Existing SDS systems are unable to enforce these policies, as they are either targeted for a specific layer (\emph{e.g.,} hypervisor, OpenStack, Ceph) and are not directly applicable over the KVS or POSIX layers~\cite{Malacology:2017:Sevilla,Crystal:2017:Gracia,IOFlow:2013:Thereska,Retro:2015:Mace}, or do not provide context propagation~\cite{Libra:2014:Shue,Pisces:2012:Shue}, thus being unable to provide differentiated treatment of I/O requests.

\input{input/floaters/table-lines-of-code}

\subsection{Per-Application Bandwidth Control}
\label{subsec:qos-diff}

The ABCI supercomputer is designed upon the convergence between AI and HPC workloads.
One of the most used AI frameworks on it is TensorFlow~\cite{TensorFlow:2016:Abadi}. 
To execute TensorFlow jobs, a user can allocate a full node, or a fraction of it where jobs can execute concurrently. 
Compute nodes are partitioned into resource-isolated \emph{instances} using Linux's cgroups~\cite{cgroups}.  
Each instance has exclusive access to CPU cores, memory space, a GPU, and local disk storage quota. 
However, local disk bandwidth is still shared.
Because each instance is agnostic of others, jobs compete for disk bandwidth leading to I/O interference and performance variation. 
Even if the block I/O scheduler can ensure fairness, all instances are served with the same service level.
This scenario prevents the possibility of assigning different priorities and achieving per-application bandwidth policies.

Using cgroups's block I/O controller (\emph{blkio}) allows static rate limiting \texttt{read} and \texttt{write} operations of each ins\-tan\-ce~\cite{BLKIO}.
However, once the I/O rate is set it cannot be dynamically changed at execution time, as it requires stopping the jobs, adjusting the parameters, and restarting the jobs, which is prohibitively expensive.
This creates a second problem where if no other job is executing in the node, the instance cannot use leftover bandwidth, leading to longer execution periods.

\paragraph{\SYS}
To address this, we use a \SYS stage that implements the mechanisms to dynamically rate limit workflows at each instance, while the control plane implements a proportional sharing algorithm to orchestrate the stage and ensure all instances meet their policies.
%
% PAIO stage
Our use case focuses on the model training phase, where each instance runs a TensorFlow job that uses a single workflow to read dataset files from the file system.
TensorFlow's \texttt{read} requests are intercepted and forwarded to the stage that contains a single channel with a \texttt{DRL} enforcement object.
Contrary to \cref{subsec:tail-latency}, \SYS does not require context propagation, as policies can be met with the \emph{request type} and \emph{size} classifiers.
As TensorFlow exposes different backend interfaces (\emph{e.g.,} POSIX, HDFS, AWS S3), we extended the POSIX file system to enforce requests at the stage. 
This was achieved by adding 22 LoC, as listed in Table~\ref{table:lines-of-code}.

\input{input/floaters/algorithm-tensorflow}

At the control plane, we implemented a max-min fair share algorithm to ensure per-application bandwidth gua\-ran\-tees, as shown in Algorithm~\ref{alg:maxmin}, which is typically used for resource fairness policies~\cite{IOFlow:2013:Thereska, Retro:2015:Mace}.
Rather than assigning the minimum I/O bandwidth to each instance, it distributes leftover bandwidth whenever it is available (\emph{left}$_{B}$). 
The algorithm uses a feedback control loop that performs the following steps.
First, the overall disk bandwidth a\-vai\-la\-ble (\emph{Max}$_B$) and bandwidth demands of each application ($demand$) are defined \emph{a priori} by the system administrator or the mechanism responsible for managing resources of different job instances~\cite{SLURM:2003:Yoo}. 
The control plane starts collecting statistics of active instances, given by $I_i$ (\texttt{1}), as well as the bandwidth generated by each TensorFlow job (collected at \texttt{/proc}). 
Then, it computes the rate limit of each active instance (\texttt{3}-\texttt{10}).
If an instance's $demand$ is less than its fair share, the control plane assigns the $demand$ (\texttt{4}-\texttt{5}), assigning its fair share otherwise (\texttt{6}).
It then distributes \emph{left}$_{B}$ between all instances (\texttt{9}-\texttt{10}).
Having computed all rates, the control plane calibrates the bucket's rate of each instance in a function of $I_i$ and $rate_i$, generating the \texttt{enf\_rule}s to be submitted to each stage (\texttt{11}).  
Finally, the control plane sleeps for \emph{loop\_interval} before beginning a new control cycle (\texttt{12}).

Existing SDS solutions that target the virtualization layer could be used for enforcing these policies under cloud-based environments~\cite{IOFlow:2013:Thereska,sRoute:2016:Stefanovici,PSLO:2016:Li}.
However, under scenarios that require bare-metal access to resources such as HPC infrastructures (\emph{e.g.,} ABCI) and bare-metal cloud servers, these solutions are unfit for ensuring such objectives. 

%% file: input/floaters/algorithm-kvs.tex
\begin{algorithm}[t]
    % \footnotesize
    \small
    \begin{algorithmic}[1]
    \caption{Tail Latency Control Algorithm}
    \label{alg:silk}
    \Require \emph{KVS}$_{B} = 200$; $min_{B} = 10$

    \State $\{Fg, Fl, L_0, L_N\} \gets collect\ ()$

    % Second phase: compute rates
    \State \emph{left}$_{B} \gets $\emph{KVS}$_{B} - Fg$
            
    \State \emph{left}$_{B} \gets max\ \{$\emph{left}$_{B}\ |\ min_{B}\}$

    \If{$Fl > 0 \wedge L_0 > 0$}
        % \State $B_{Fl} \gets $\emph{left}$_{B} / 2; \ B_{L_0} \gets $\emph{left}$_{B} / 2; \ B_{L_N} \gets 0$
        \State $\{B_{Fl},\ B_{L_0},\ B_{L_N}\} \gets \{$\emph{left}$_{B} / 2,\ $\emph{left}$_{B} / 2,\  min_{B}\}$

    \ElsIf{$Fl > 0 \wedge L_0 = 0$}
        % \State $B_{Fl} \gets $\emph{left}$_{B}; \ B_{L_0} \gets 0; \ B_{L_N} \gets 0$
        \State $\{B_{Fl},\ B_{L_0},\ B_{L_N}\} \gets \{$\emph{left}$_{B},\ min_{B},\  min_{B}\}$        
    
    \ElsIf{$Fl = 0 \wedge L_0 > 0$}
        % \State $B_{L_0} \gets $\emph{left}$_{B}; \ B_{Fl} \gets 0; \ B_{L_N} \gets 0$
        \State $\{B_{Fl},\ B_{L_0},\ B_{L_N}\} \gets \{\ min_{B},\ $\emph{left}$_{B},\ min_{B}\}$
    
    \Else
        % \State $B_{L_N} \gets $\emph{left}$_{B}; \ B_{Fl} \gets 0; \ B_{L_0} \gets 0$
        \State $\{B_{Fl},\ B_{L_0},\ B_{L_N}\} \gets \{\ min_{B},\ min_{B},\ $\emph{left}$_{B}\}$
    
    \EndIf

    % \State $\emptyset \gets$ \emph{enforce (\{$B_{Fl},\ B_{L_0},\ B_{L_N}$\})}
    \State \emph{enf\_rule (\{$B_{Fl},\ B_{L_0},\ B_{L_N}$\})}

    \State \emph{sleep ($loop\_interval$)}

    \end{algorithmic}
\end{algorithm}

%% file: input/floaters/table-lines-of-code.tex
\begin{table}[t]
    \centering
    \caption{\emph{Lines of code added to RocksDB and TensorFlow.}}
    \vspace*{-5pt}
    \small
    \begin{tabular}{rcc}
        \toprule
        \multirow{2}{*}{}                   & \multicolumn{2}{c}{\small\textbf{Lines added}}                \\
                                            &  {\small RocksDB$^\dagger$} & {\small TensorFlow$^\dagger$}   \\ \cline{2-3} 
        Initialize \SYS stage               &  10                   & 15                                    \\
        Context propagation instr.          &  47                   & --                                    \\
        Serialize \emph{context} object     &  7                    & 3                                     \\
        Instrument operation calls          &  17                   & 2                                     \\
        Deserialize \emph{result} object    &  4                    & 2                                     \\ \midrule
        {\small\textbf{Total}}              &  85                   & 22                                    \\ \bottomrule
        \multicolumn{3}{l}{\scriptsize$^\dagger$RocksDB~\cite{rocksdb-git:2020} and TensorFlow~\cite{tensorflow-git:2020} codebases consist of approximately}  \\
        \multicolumn{3}{l}{\scriptsize 335K LoC and 2.3M LoC, respectively.} \\
    \end{tabular}
    \label{table:lines-of-code}
    \vspace*{-5pt}
\end{table}

%% file: input/floaters/algorithm-tensorflow.tex
\begin{algorithm}[t]
    % \footnotesize
    \small
    \begin{algorithmic}[1]
    \caption{Max-min Fair Share Control Algorithm}
    \label{alg:maxmin}
    \Require \emph{Max}$_{B} = 1GiB$;  \emph{Active} $> 0$; \emph{demand}$_i > 0$
    % \Ensure Assign \emph{$R_i$} to \emph{$Stage_i$}; ... 

    \State $\{I_1,\ I_2,\ I_3,\ I_4\} \gets collect\ ()$
    
    % Second phase: compute rates
    \State \emph{left}$_{B} \gets $\emph{Max}$_B$
    
    \For{$i=0$ in [0, \emph{Active}$ - 1$]}
        \If{\emph{demand}$_i \leq \frac{left_B}{Active - i}$}
            \State \emph{rate}$_i \gets$ \emph{demand}$_i$ 
        \Else 
            \State \emph{rate}$_i \gets \frac{left_B}{Active - i}$
        \EndIf
        \State \emph{left}$_{B} \gets$ \emph{left}$_{B}$ - \emph{rate}$_i$ 
    \EndFor

    \For{$i=0$ in [0, \emph{Active}$ - 1$]}
        \State \emph{rate}$_i \gets \frac{left_B}{Active}$
    \EndFor

    \State \emph{enf\_rule $(\{$rate$_1, I_1\},\{$rate$_2, I_2\},\{$rate$_3, I_3\},\{$rate$_4, I_4\})$}
    % \State \emph{enforce $(\{$\emph{rate}$_1, \{I_1,\ I_2,\ I_3,\ I_4\})$}

    \State \emph{sleep ($loop\_interval$)}

    \end{algorithmic}
\end{algorithm}

%% file: input/06-evaluation.tex
\section{Evaluation}
\label{sec:evaluation}

Our evaluation seeks to demonstrate the \emph{performance and scalability} of \SYS (\cref{subsec:paio-performance}), and its \emph{ability and feasibility of enforcing I/O policies} over different scenarios (\cref{subsec:eval-tail-latency} -- \cref{subsec:eval-qos-diff}).
Unless stated otherwise, experiments were conducted in a compute node of the ABCI supercomputer with the following hardware specifications: two 20-core Intel Xeon processors with 2-way multi-threading, four NVidia Tesla V100 GPUs, 384GiB of RAM, and a 1.6TiB Intel SSD DC P4600.
It uses CentOS 7.5 with Linux kernel 3.10 and the \texttt{xfs} file system.

\subsection{\SYS Performance and Scalability}
\label{subsec:paio-performance}

\begin{figure}
    \includegraphics[width=1\linewidth,keepaspectratio]{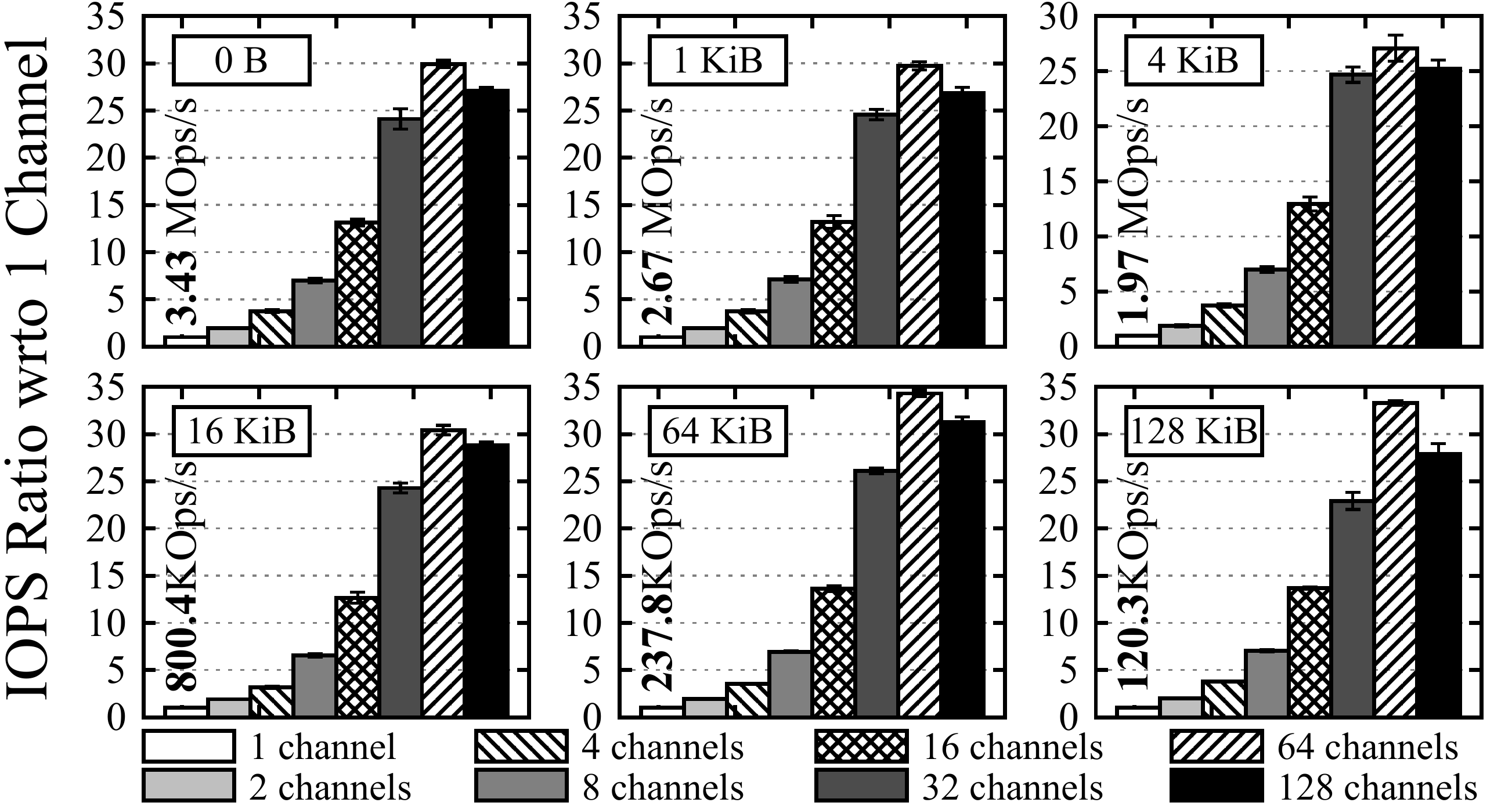}%}
    \vspace*{-5pt}
    \caption{\textbf{Cumulative IOPS} \emph{of} \SYS \emph{channels (1--128) un\-der different request sizes (0 --128 KiB) with respect to 1 channel.}}
    \label{fig:microbench-iops}
\end{figure}

We developed a simple benchmark that simulates an application whose requests are enforced with a \SYS stage.
This benchmark aims to demonstrate the maximum performance achievable with \SYS by stress-testing it in a loop-back manner. 
It generates and submits multi-threaded requests in a closed loop through \emph{Instance}'s \texttt{enforce} call, under a varying number of clients (\emph{e.g.,} workflows) and request sizes.
Request size and number of client threads range between 0--128KiB and 1--128, respectively.
A \SYS stage is configured with varying number of channels (matching the number of client threads), each containing a \texttt{Noop} enforcement object that copies the request's buffer to the \emph{result} object.
All reported results are the mean of at least ten runs, and standard deviation is kept below 5\%.
Experiments were conducted using a machine with two 18-core Intel Xeon processors with 2-way multi threading, configured with Ubuntu Server 20.04 and Linux kernel 5.8.9.

\input{input/06-evaluation-usecase1-plots}

\paragraph{IOPS and Bandwidth}
Fig.~\ref{fig:microbench-iops} depicts the cumulative IOPS ratio with respect to a single channel.
Absolute IOPS value is shown above the \emph{1 channel} bar. 
A 0B request size represents a \emph{context}-only request (\emph{i.e.,} no content), as described in~\cref{subsec:enforcement}.

When using a 0B request size, a single \SYS channel achieves an average throughput of 3.43 MOps/s.
Since the workload is CPU-bound, client threads start competing for processing time, and thus, \SYS achieves higher throughput when using 64 channels.
Under this configuration, \SYS achieves a cumulative throughput of 102.7 MOps/s, representing a 30$\times$ performance increase.

As the request size increases so does the total bytes processed by \SYS. 
When configured with 64 channels, \SYS is able to process 489 GiB/s for 128KiB-sized requests.
As for a single channel, \SYS processes requests at 2.5 GiB/s and 14.7 GiB/s for 1KiB and 128KiB request sizes, respectively.

\paragraph{Profiling}
We measured the execution time of each \SYS operation that appears in the main execution path.
\emph{Context} object creation takes approximately 17~ns, while the channel and enforcement object selection take 85~ns to complete (each).
The duration of \texttt{obj\_enf} ranges between 20~ns and 7.45~$\mu$s when configured with 0B and 128KiB request sizes.

\vspace{.2\baselineskip}\noindent\textbf{Summary:}
This section showcases the maximum performance achieved with \SYS when enforcing \emph{context}-only operations (\emph{i.e.,} 0B-sized requests) and requests with varying sizes.
Throughput-wise, \SYS achieves, at most, 102.7 MOps/s and is able to process I/O requests at 489 GiB/s.
Latency-wise, a single \SYS channel enforces requests at 291 ns and 8.31 $\mu$s, for 0B and 128KiB request sizes, respectively.

\subsection{Tail Latency Control in Key-Value Stores}
\label{subsec:eval-tail-latency}

We now demonstrate how \SYS achieves tail latency control under several workloads and how does it compare to other systems.
We compare the performance of RocksDB with Auto-tuned (\emph{i.e.,} a version of RocksDB with auto-tuned rate limiting of background operations~\cite{AutoTunedRateLimiter:2020}), SILK, and \SYS (\emph{i.e.,} a \SYS-enabled RocksDB).
We tuned all KVS with the following configurations.
The \texttt{memtable-size} was set to 128MiB. 
We used 8 worker threads for client operations and 8 background threads for flush (1) and compactions (7).
The minimum bandwidth threshold for background operations was set to 10MiB/s.
To simplify results compression and commit logging are turned off.
All experiments were conducted using the \texttt{db\_bench} benchmark suite~\cite{dbbench:2020}, and resources were limited using Linux cgroups~\cite{cgroups,BLKIO}.
We limit memory usage to 1GiB and I/O bandwidth to 200MiB/s, as used in the SILK testbed, which is based on Nutanix production environments~\cite{SILK:2019:Balmau}.
Conducting ex\-pe\-ri\-ments with higher limits would lead to similar results, however it would require longer execution periods and a larger dataset to generate a similar backlog.

We focus on workloads made of bursty clients, to better simulate existing services in production~\cite{SILK:2019:Balmau,RocksDBFacebook:2020:Cao}.
Client requests are issued in a closed loop through a combination of peaks and valleys.
An initial valley of 300 seconds submits operations at 5kops/s, and is used for executing the KVS internal backlog.
Peaks are issued at a rate of 20kops/s for 100 seconds, followed by 10 seconds valleys at 5kops/s.
All datastores were preloaded with 100M key-value pairs, using a uniform key-distribution, 8B keys and 1024B values.

We use three workloads with different read:write ratios, namely \emph{mixture} with 50:50, \emph{read-heavy} with 90:10, and \emph{write-heavy} with 10:90.
\emph{Mixture} represents a commonly used YCSB workload (\emph{i.e.,} workload A) and provides a similar ratio as Nutanix production workloads~\cite{SILK:2019:Balmau}.
\emph{Read-heavy} provides an operation ratio similar to those reported at Facebook~\cite{RocksDBFacebook:2020:Cao}.
To present a comprehensive testbed, we included a \emph{write-heavy} workload.
For each system, workloads were executed three times over 1-hour with a uniform key-dis\-tri\-bu\-ti\-on. 
For figure clarity, we present the first 20 minutes of a single run.
Similar performance curves were observed for the rest of the execution.
Fig.~\ref{fig:usecase1-mixture}--\ref{fig:usecase1-writeheavy} depict throughput and $99^{th}$ percentile latency of all systems under each workload. 
Theoretical client load is presented as a red dashed line. 
Mean throughput is shown as an horizontal dashed line.

\paragraph{Mixture workload}
Fig.~\ref{fig:usecase1-mixture} depicts the results of each system under the mixture workload.
Due to accumulated backlog of the loading phase, throughput achieved in all systems does not match the theoretical client load. 
RocksDB presents high tail latency spikes due to constant flushes and low level compactions.
Auto-tuned presents less latency spikes but degrades overall throughput, which occurs due to the rate limiter being agnostic of background tasks' priority, and because it increases its rate when there is more backlog, contending for disk bandwidth.
SILK achieves low tail latency but suffers periodic drops in throughput due to accumulated backlog.
Compared to RocksDB (11.9~kops/s), \SYS provides similar mean throughput (12.4~kops/s).
Regarding tail latency, while RocksDB experiences peaks that range between 3--20~ms, \SYS and SILK observe a 4x decrease in absolute tail latency, with values ranging between 2--6~ms.

\paragraph{Read-heavy workload}
Fig.~\ref{fig:usecase1-readheavy} depicts the results under the read-heavy workload.
Throughput-wise all systems perform identically.
At different periods, all systems demonstrate a temporary throughput degradation due to accumulated backlog.
As for tail latency, the analysis is twofold.
RocksDB and Auto-tuned present high tail latency up to the 400~s mark.
After that mark, RocksDB does not have more pending backlog and achieves sustained tail latency that ranges between 1--3~ms, while on Auto-tuned, some compactions are still being performed due to rate limiting, thus increasing latency by 1 to 2~ms. 
SILK and \SYS have similar tail latency curves.
During the initial valley both systems significantly improve tail latency when compared to RocksDB.
After the 400~s mark, SILK pauses high level compactions and presents a tail latency between 1--2~ms. 
By preempting high level compactions and serving low level ones through the same thread pool as flushes, it ensures that high priority tasks are rarely stalled.
SILK achieves this by modifying the original RocksDB queuing mechanism. 
In \SYS, while sustained, its tail latency is 1~ms higher than SILK's in the same observation period.
Since \SYS does not modify the RocksDB engine, it cannot preempt compactions, resulting in a small increase on client latency.

\begin{figure*}[t]
    \centering
    \includegraphics[width=1\textwidth, height=79pt]{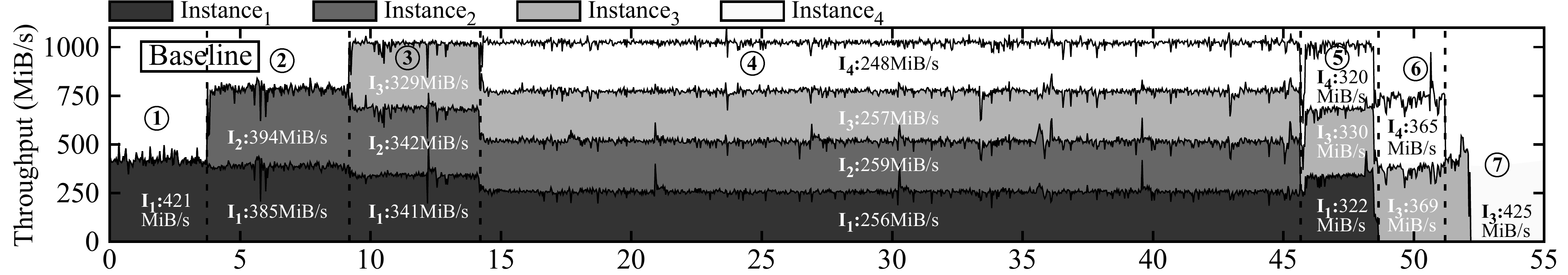} \\
    \vspace*{2.5pt}
    \includegraphics[width=1\textwidth, height=73pt]{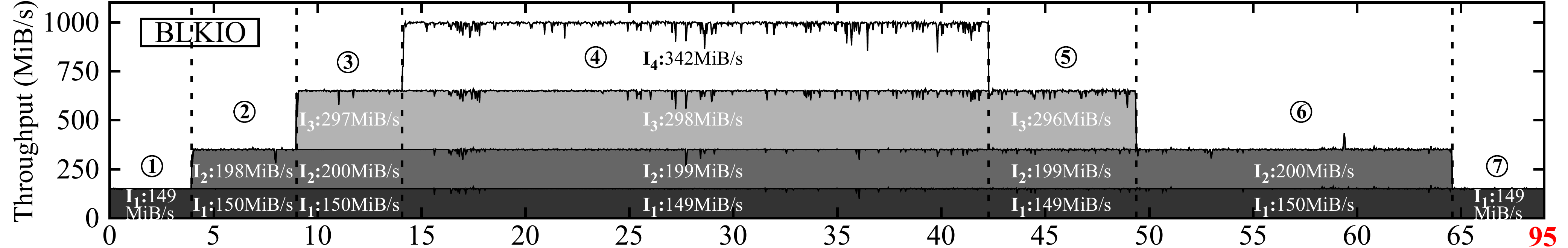} \\ 
    \vspace*{2.5pt}
    \includegraphics[width=1\textwidth, height=79pt]{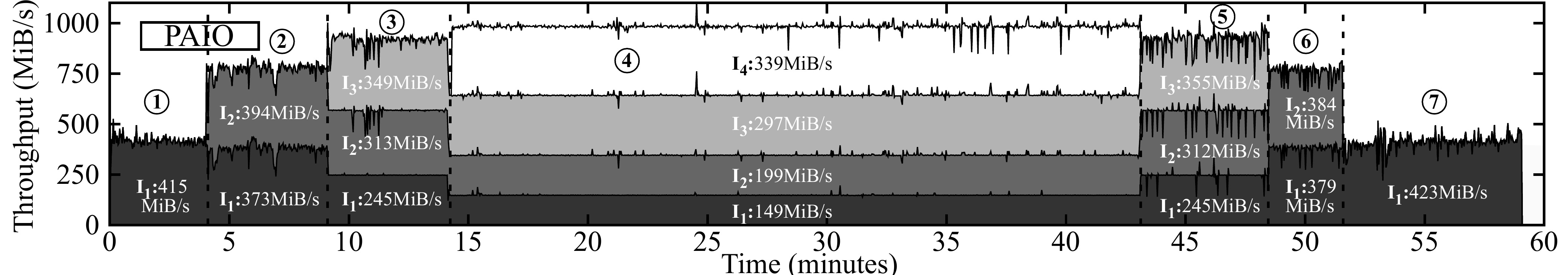} \\

    \vspace*{-5pt}
    \caption{   
        \textbf{Per-application bandwidth limits under shared storage} \emph{for Baseline, Blkio, and} \SYS \emph{setups.} 
    }
    \label{fig:usecase2}
    \vspace*{-10pt}
\end{figure*}

\paragraph{Write-heavy workload}
Fig.~\ref{fig:usecase1-writeheavy} depicts the results under the write-heavy workload.
As high write proportions continuously generate latency-critical background tasks, RocksDB is not able to endure this load, resulting in high latency spikes.
Auto-tuned limits all background writes, which degrades latency spikes, but still achieves 5~ms tail latencies in several periods. 
SILK pauses all high level compactions and only latency-critical tasks are served, improving mean throughput and keeping latency spikes below the 5~ms mark.
In \SYS, since flushes occur more frequently, the control plane slows down high level compactions more aggressively, which leads to low level ones to be temporary halted at the compaction queue, waiting to be executed. 
While degrading mean throughput, \SYS still decreases tail latency, never exceeding 7~ms. 
The throughput difference between \SYS and SILK is justified by the latter preempting high level compactions over low level ones, as described in the read-heavy workload.

\paragraph{Takeaway}
We demonstrate that by abstracting a minimal amount of the application's semantics (\emph{i.e.,} context) and propagating it to the data plane stage with minor changes to complex codebases (\emph{i.e.,} 47 LoC), \SYS outperforms RocksDB, an industry-standard KVS, by at most 4$\times$ in tail latency, and enables as much control and performance as system-specific optimizations (SILK) that required profound refactoring to the original codebase.

\subsection{Per-Application Bandwidth Control}
\label{subsec:eval-qos-diff}

We now show how \SYS ensures per-application bandwidth guarantees under a shared storage environment.
Our setup was driven by the requirements of the ABCI supercomputer.
Experiments ran using TensorFlow 2.1.0 with the LeNet training model, configured with a batch size of 64 TFrecords.
We used the Imagenet dataset. 
Each instance runs with a dedicated GPU and dataset, and its memory is limited to 32GiB.
Overall disk bandwidth is rate limited to 1GiB/s. 

At all times, a node executes at most four instances with equal resource shares in terms of CPU, GPU, and RAM. 
Each instance executes a TensorFlow job, is assigned with a bandwidth policy, and executes a given number of training epochs.
Namely, instances \textbf{1} to \textbf{4} are assigned with minimum bandwidth guarantees of \textbf{150}, \textbf{200}, \textbf{300}, and \textbf{350} MiB/s, and execute \textbf{6}, \textbf{5}, \textbf{5}, and \textbf{4} training epochs, respectively.

Experiments were conducted under three setups. 
\emph{Baseline} represents the current setup supported at the ABCI supercomputer, where all instances execute without bandwidth guarantees.
In \emph{Blkio}, the bandwidth requirements are defined using blkio~\cite{BLKIO}.
In \SYS, each instance executes with a \SYS stage that enforces the specified bandwidth goals while dynamically distributing leftover bandwidth.

Fig.~\ref{fig:usecase2} illustrates, for each setting, the I/O bandwidth at 1-second intervals. 
Experiments include seven phases, each marking when an instance starts or completes its execution.

\paragraph{Baseline}
Experiments were executed over 52 minutes. 
At \numlightcirclesans{1}, instance $I_1$ reads at 421~MiB/s.
Whenever a new instance is added, the I/O bandwidth is shared across all (\numlightcirclesans{2}). 
At \numlightcirclesans{3}, the aggregate instance throughput matches the disk limit. 
At \numlightcirclesans{4}, instance performance converges to approximately 256~MiB/s, leading to all instances experiencing the same service level.
However, $I_3$ and $I_4$ cannot meet their goal, since $I_1$ and $I_2$ have more than their fair share. 
After 46 minutes of execution (\numlightcirclesans{5}), $I_3$ terminates, and leftover bandwidth is shared with the remainder.
Again, $I_4$ cannot achieve the targeted service level.
At \numlightcirclesans{6} and \numlightcirclesans{7}, active instances have access to leftover bandwidth and finish their execution.

\noindent\textbf{Summary:} 
Instances $I_3$ and $I_4$ were unable to achieve their bandwidth guarantees, missing their deadlines during 31 and 34 minutes, respectively.

\paragraph{Blkio}
Experiments were executed over 95 minutes.
During the execution (\numlightcirclesans{1} to \numlightcirclesans{7}), whenever a new instance is added, it is provisioned with its bandwidth limit.
However, because the rate of each instance is set using blkio, instances cannot use leftover bandwidth to speed up their execution. 
For example, while on \emph{Baseline} $I_1$ executes under the 50-minutes mark, it takes 95 minutes to complete its execution in \emph{Blkio}.
To overcome this, a possible solution would require to stop and checkpoint the instance's execution, reconfigure blkio with a new rate, and resume from the latest checkpoint.
However, doing this process every time a new instance joins or leaves the compute node would significantly delay the execution time of all running instances.

\noindent\textbf{Summary:} 
All instances achieve their bandwidth guarantees but cannot be dynamically provisioned with available disk bandwidth, leading to longer periods of execution.

\paragraph{\SYS}
Experiments were executed over 59 minutes.
At \numlightcirclesans{1} and \numlightcirclesans{2}, instances are assigned with their proportional I/O share, as the control plane first meets each instance demands and then distributes leftover bandwidth proportionally.
Contrary to \emph{Baseline}, where all active instances experience the same service level, at \numlightcirclesans{3}, the control plane bounds the bandwidth of $I_1$ and $I_2$ to a mean throughput of 245~MiB/s and 313~MiB/s, respectively.
At \numlightcirclesans{4}, instances are set with their bandwidth limit.
During this phase, \SYS provides the same properties as blkio.
From \numlightcirclesans{5} to \numlightcirclesans{7}, as instances end their execution, active ones are provisioned as in \numlightcirclesans{1} to \numlightcirclesans{3}.

\noindent\textbf{Summary:}
All instances met their bandwidth objectives.
When all instances are active, \SYS matches the performance of \emph{Blkio}. 
When leftover bandwidth is available, \SYS shares it across all active instances, speeding up their execution.
Compared to \emph{Blkio}, \SYS finishes 36, 13, and 2 minutes faster for $I_1$, $I_2$, and $I_3$, and performs identically for $I_4$.

\paragraph{Takeaway}
We demonstrate that \SYS can ensure per-application bandwidth guarantees under real shared storage environments.
Compared to \emph{Baseline}, which represents the current setup at the ABCI supercomputer, \SYS ensures that policies are met at all times. 
Compared to \emph{Blkio}, as \SYS distributes leftover bandwidth proportionally across active instances, it significantly reduces the overall execution time.

%% file: input/06-evaluation-usecase1-plots.tex
\begin{figure*}[t]
    \centering
    \subfloat{%
        \includegraphics[width=.50\linewidth, height=37pt]{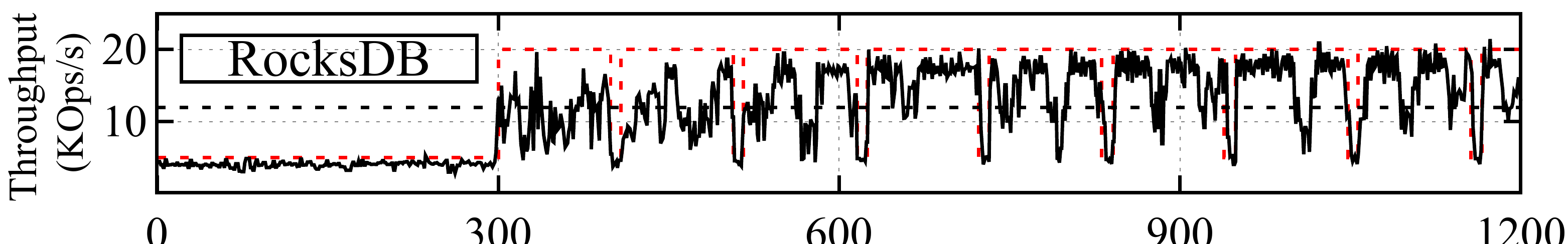}}
    \subfloat{%
        \includegraphics[width=.50\linewidth, height=37pt]{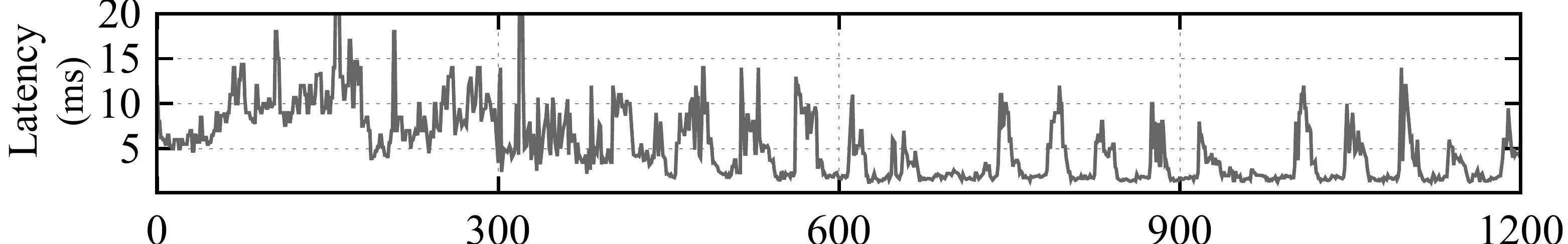}} \\
    \vspace*{-16.5pt}
    \subfloat{%
        \includegraphics[width=.50\linewidth, height=37pt]{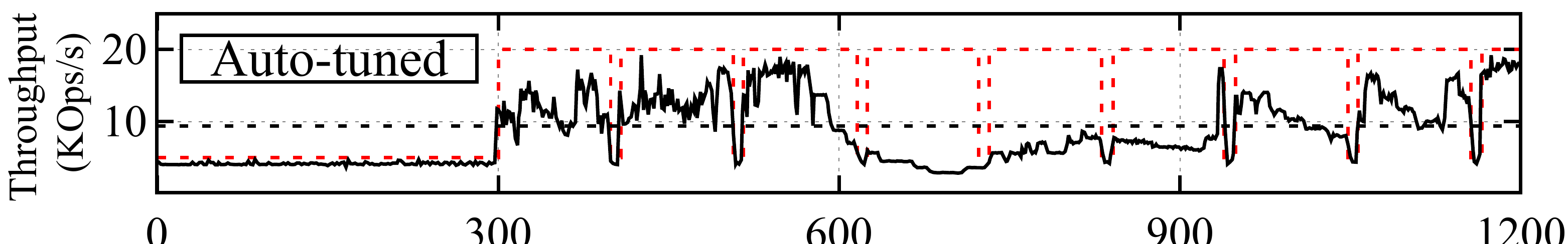}}
    \subfloat{%
        \includegraphics[width=.50\linewidth, height=37pt]{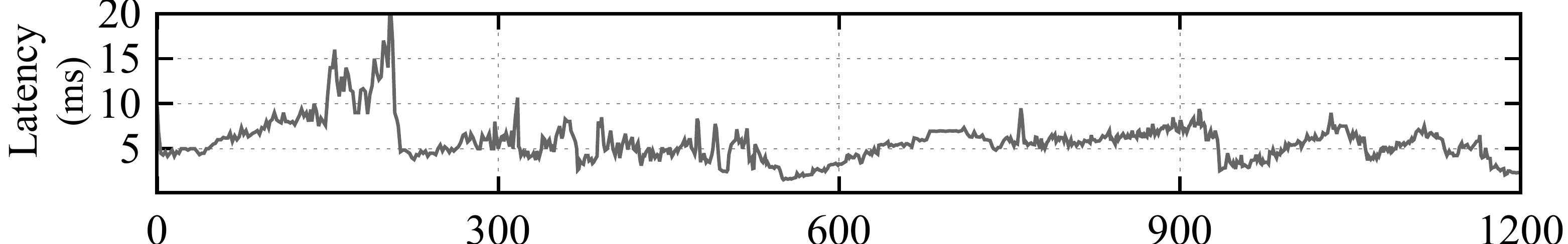}} \\
    \vspace*{-16.5pt}
    \subfloat{%
        \includegraphics[width=.50\linewidth, height=37pt]{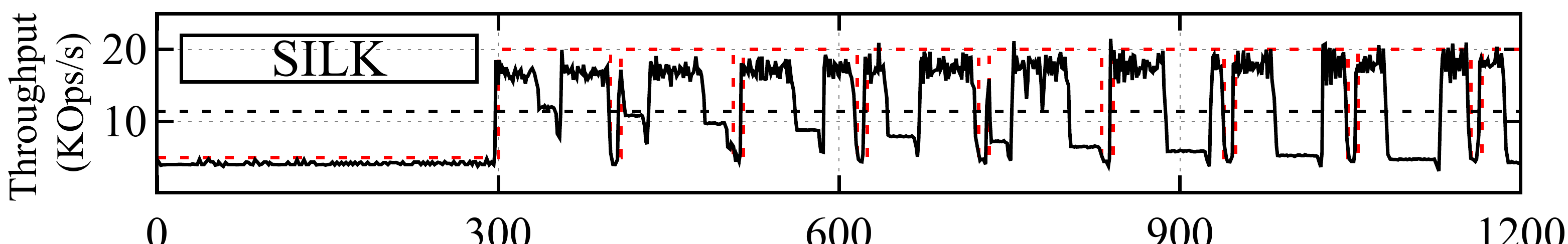}}
    \subfloat{%
        \includegraphics[width=.50\linewidth, height=37pt]{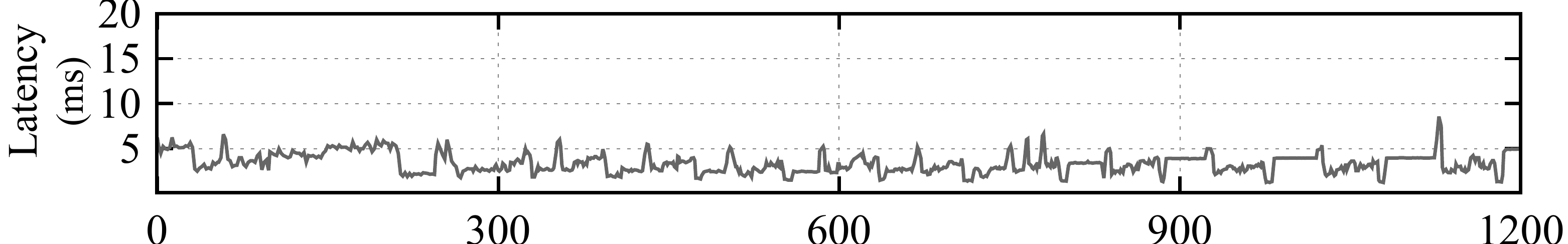}} \\
    \vspace*{-16.5pt}
    \subfloat{%
        \includegraphics[width=.50\linewidth, height=42pt]{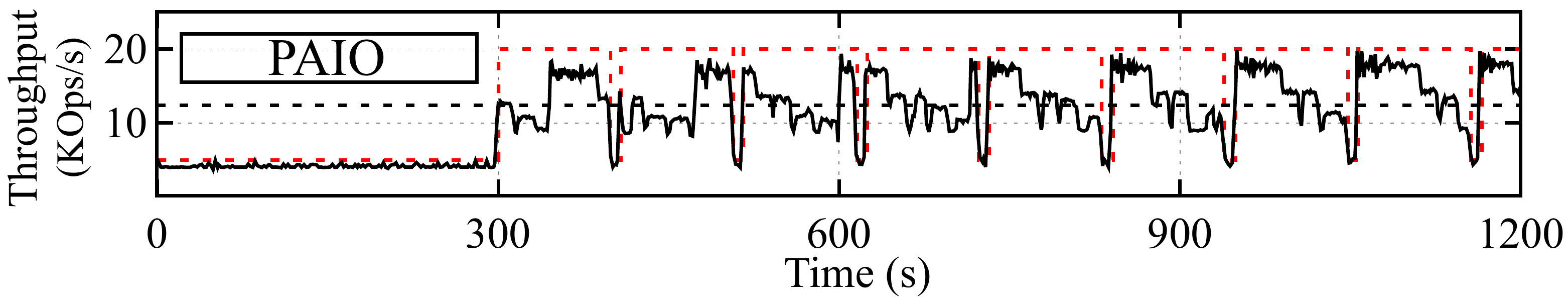}}
    \subfloat{%
        \includegraphics[width=.50\linewidth, height=42pt]{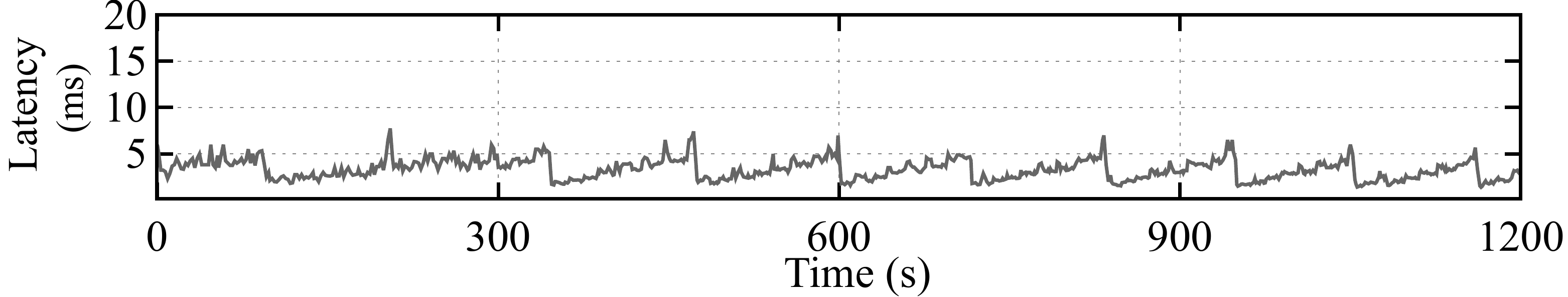}} \\

    \vspace*{-9pt}
    \caption{   
        \textbf{Mixture workload.} 
        \emph{Throughput and $99^{th}$ percentile latency results for RocksDB, Auto-tuned, SILK, and} \SYS.
    }
    \label{fig:usecase1-mixture}
    \vspace*{-23pt}
\end{figure*}

\begin{figure*}[!h]
\centering
    \subfloat{%
        \includegraphics[width=.50\linewidth, height=37pt]{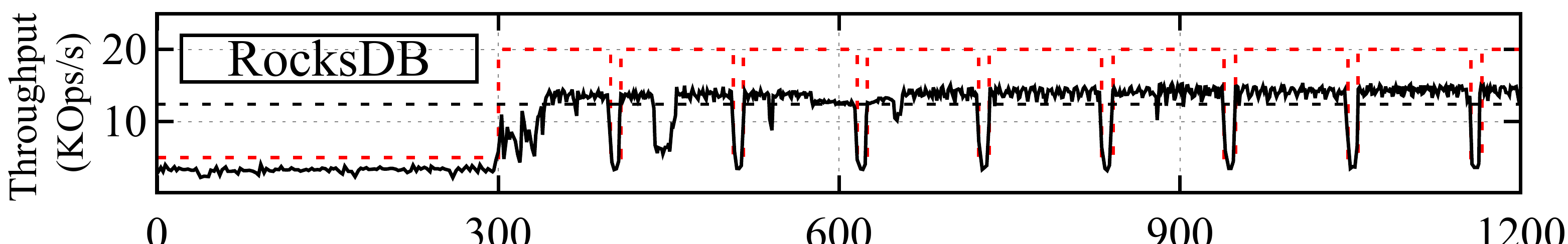}}
    \subfloat{%
        \includegraphics[width=.50\linewidth, height=37pt]{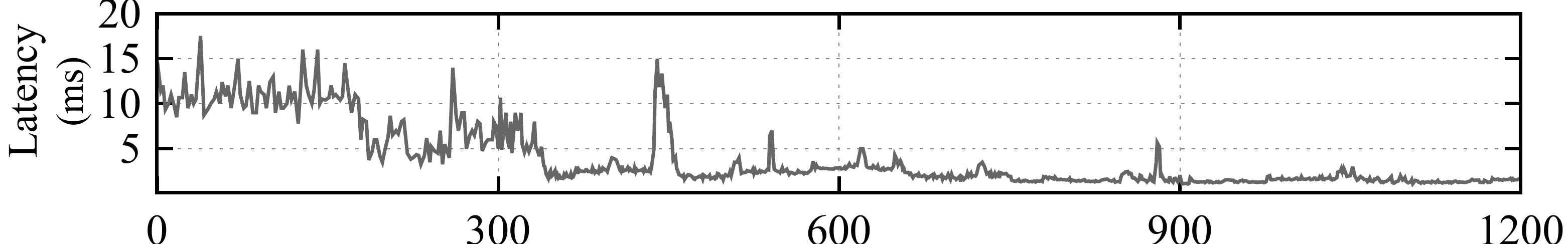}} \\
    \vspace*{-16.5pt}
    \subfloat{%
        \includegraphics[width=.50\linewidth, height=37pt]{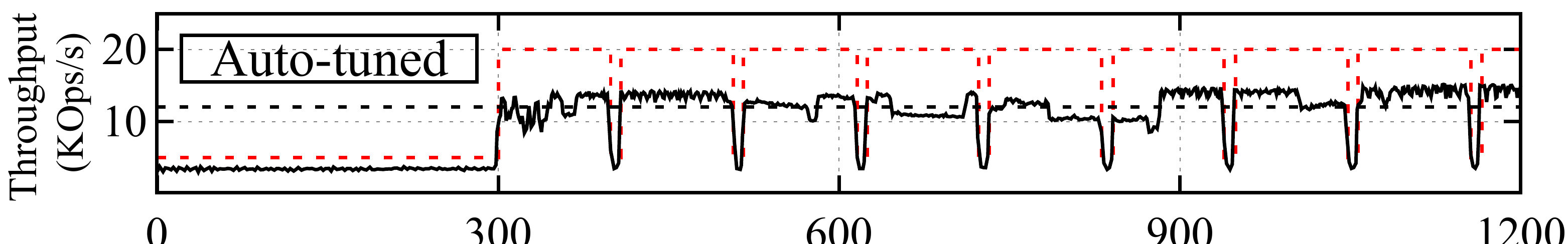}}
    \subfloat{%
        \includegraphics[width=.50\linewidth, height=37pt]{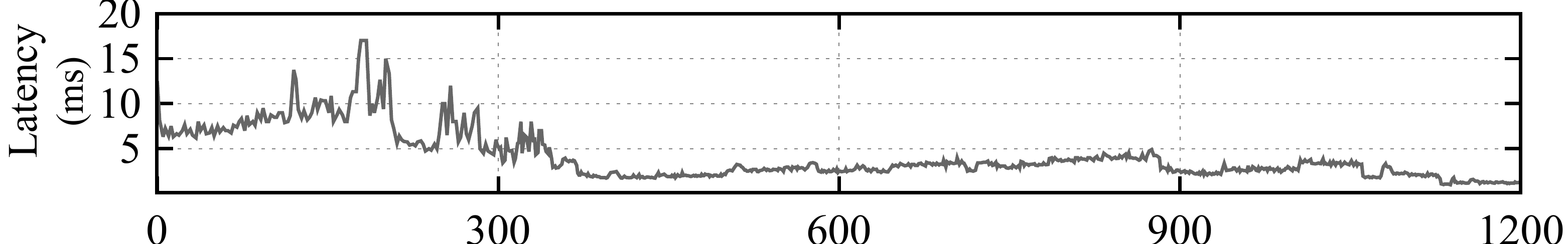}} \\
    \vspace*{-16.5pt}
    \subfloat{%
        \includegraphics[width=.50\linewidth, height=37pt]{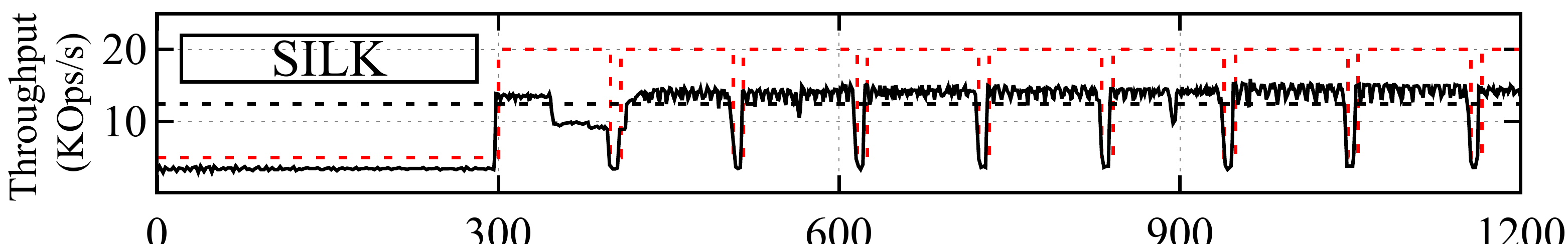}}
    \subfloat{%
        \includegraphics[width=.50\linewidth, height=37pt]{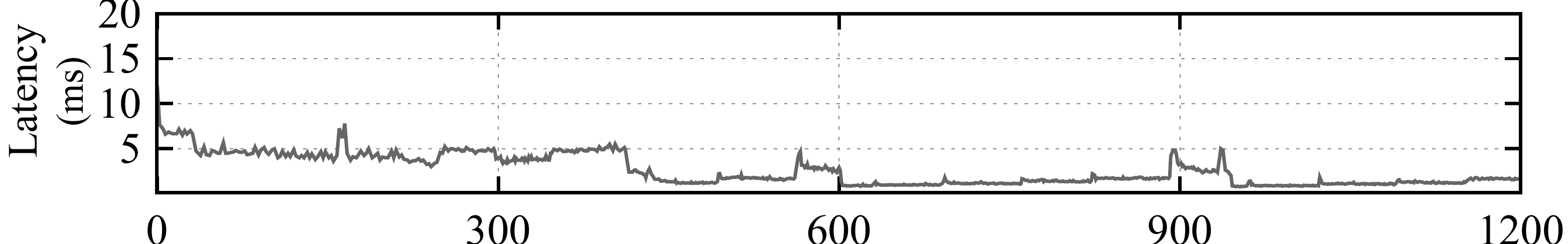}} \\
    \vspace*{-16.5pt}
    \subfloat{%
        \includegraphics[width=.50\linewidth, height=42pt]{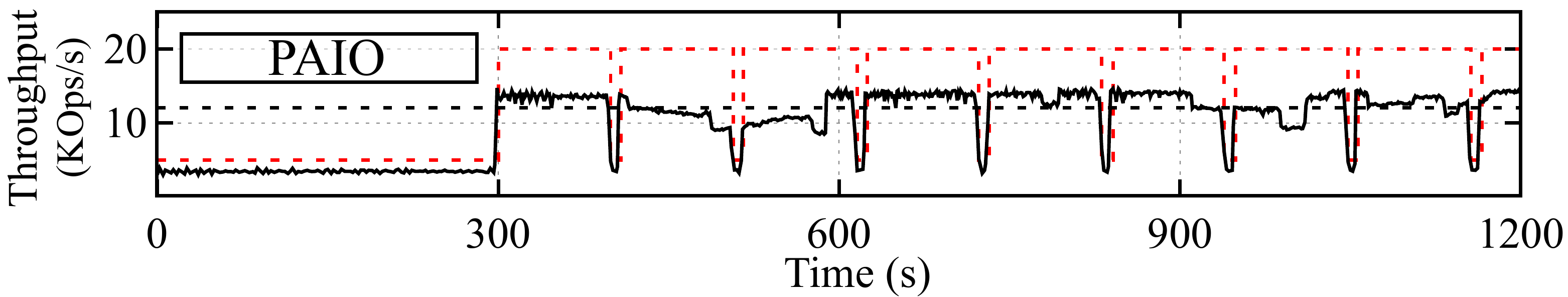}}
    \subfloat{%
        \includegraphics[width=.50\linewidth, height=42pt]{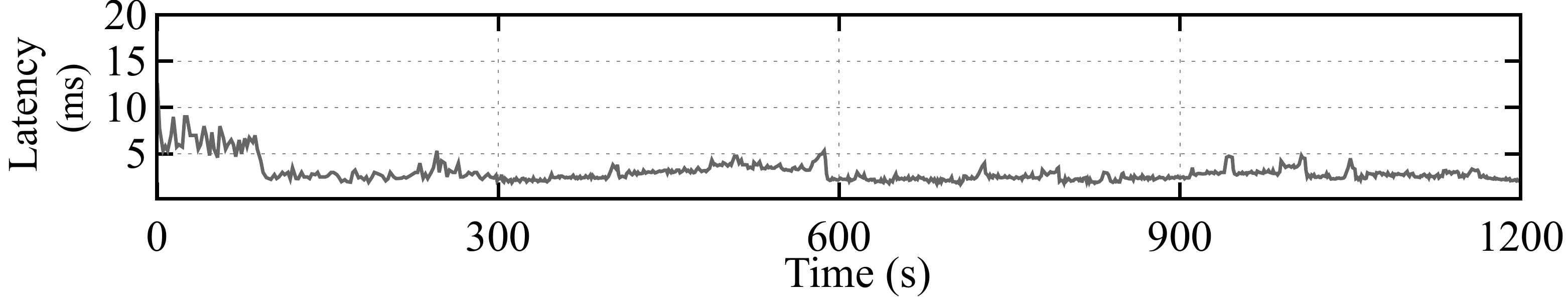}} \\

    \vspace*{-9pt}
    \caption{   
        \textbf{Read-heavy workload.} 
        \emph{Throughput and $99^{th}$ percentile latency results for RocksDB, Auto-tuned, SILK, and} \SYS.
    }
    \label{fig:usecase1-readheavy}
    \vspace*{-23pt}
\end{figure*}

\begin{figure*}[!h]
    \centering
    \subfloat{%
        \includegraphics[width=.50\linewidth, height=37pt]{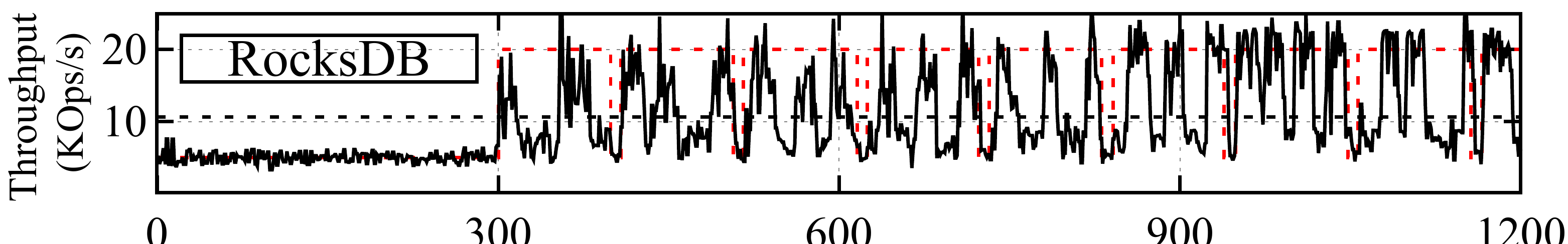}}
    \subfloat{%
        \includegraphics[width=.50\linewidth, height=37pt]{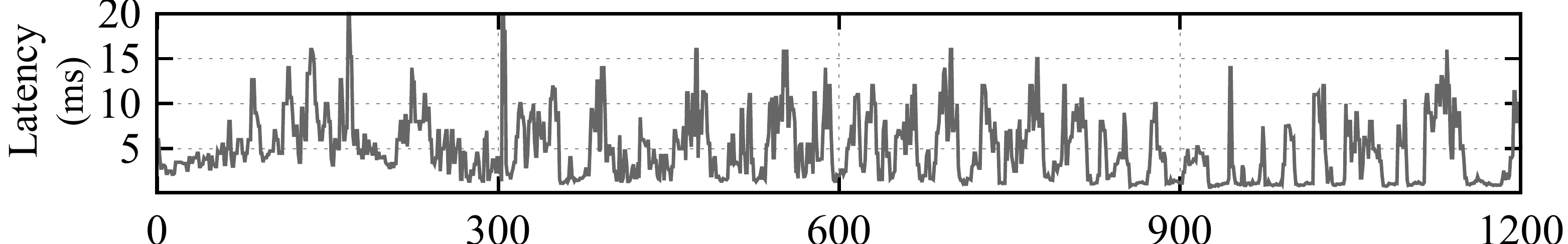}} \\
    \vspace*{-16.5pt}
    \subfloat{%
        \includegraphics[width=.50\linewidth, height=37pt]{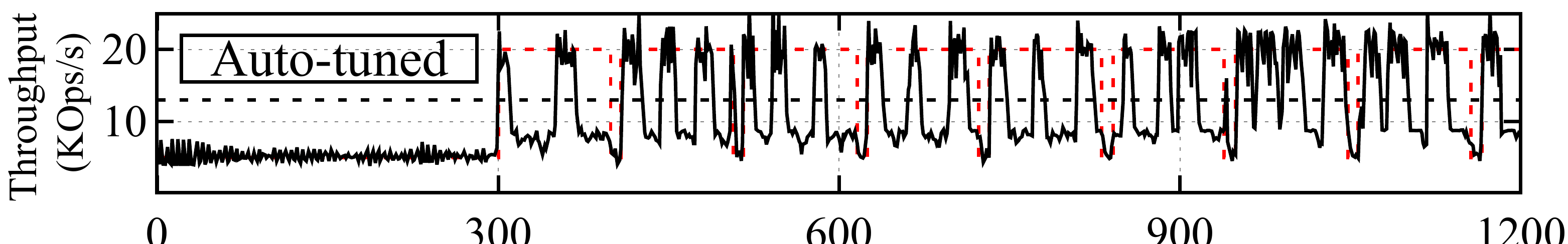}}
    \subfloat{%
        \includegraphics[width=.50\linewidth, height=37pt]{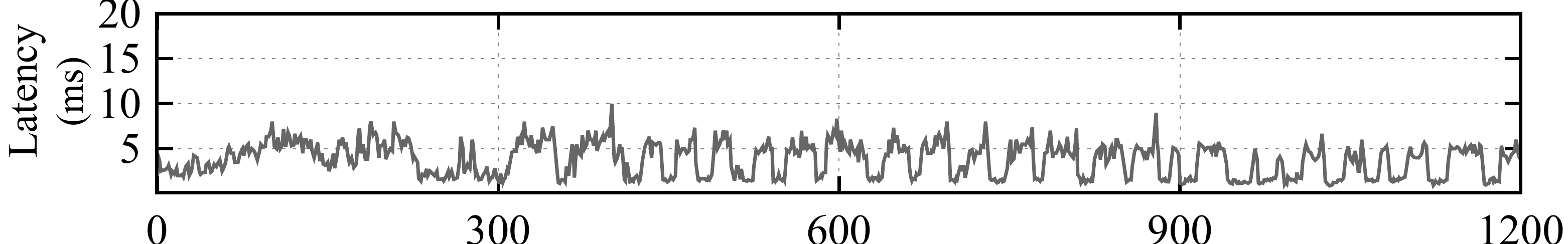}} \\
    \vspace*{-16.5pt}
    \subfloat{%
        \includegraphics[width=.50\linewidth, height=37pt]{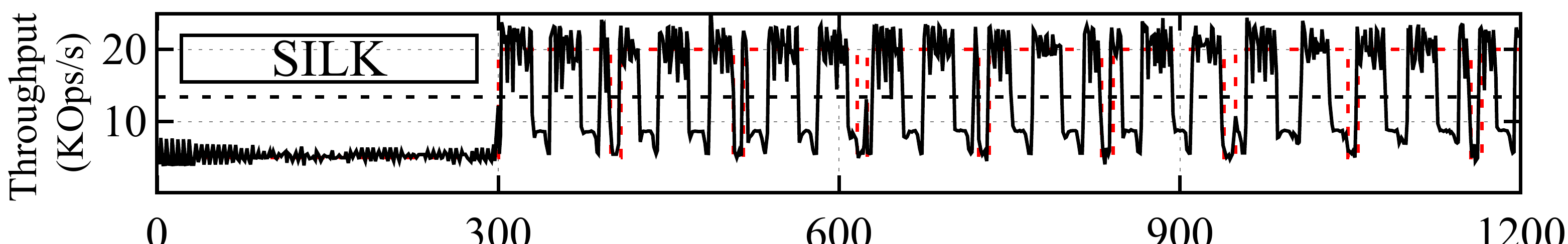}}
    \subfloat{%
        \includegraphics[width=.50\linewidth, height=37pt]{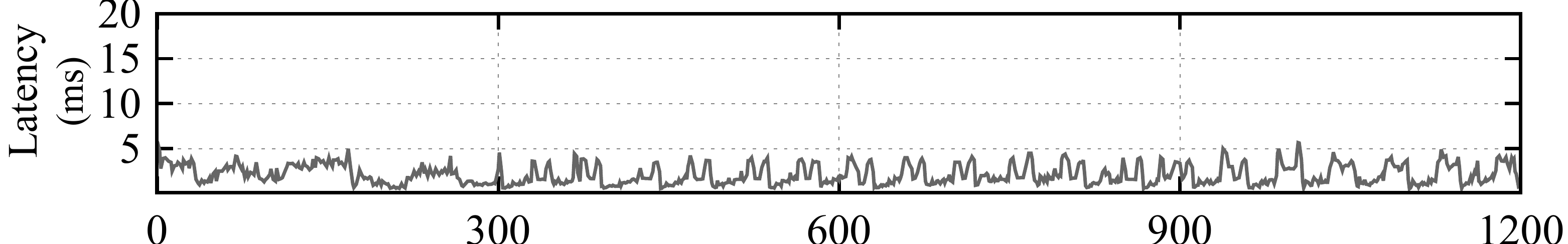}} \\
    \vspace*{-16.5pt}
    \subfloat{%
        \includegraphics[width=.50\linewidth, height=42pt]{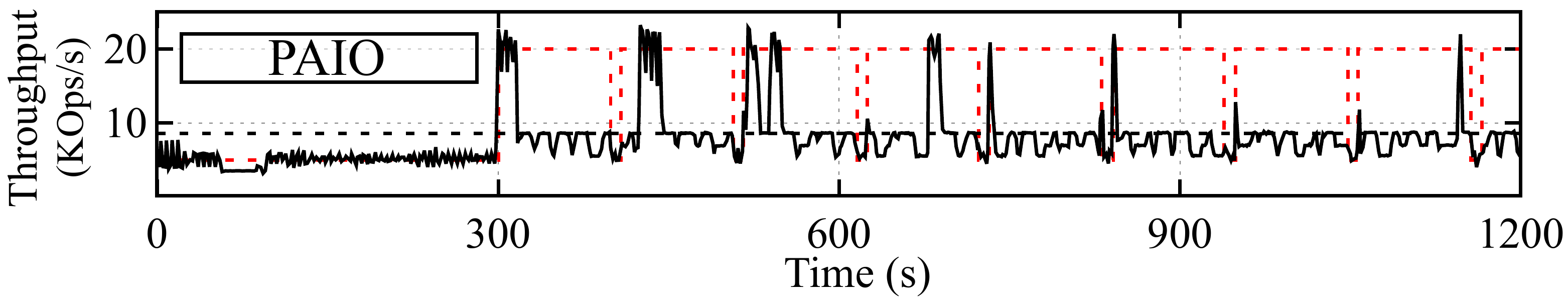}}
    \subfloat{%
        \includegraphics[width=.50\linewidth, height=42pt]{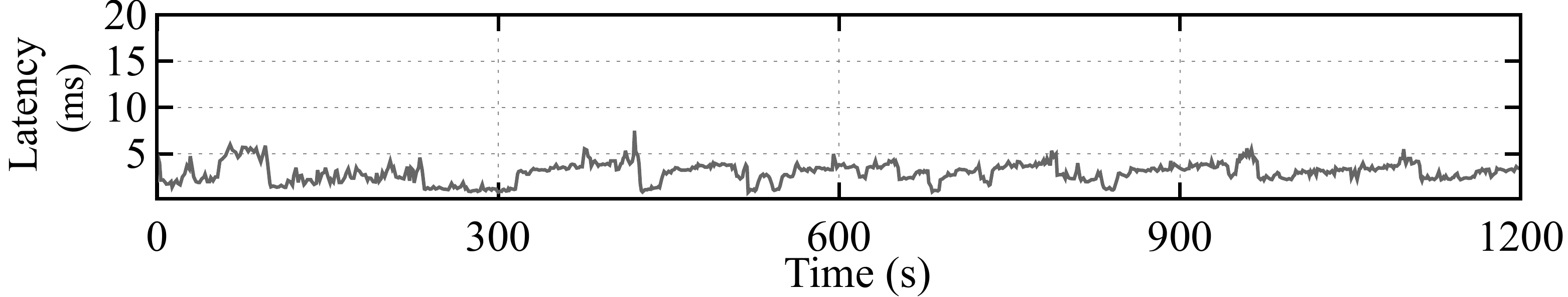}} \\

    \vspace*{-9pt}
    \caption{   
        \textbf{Write-heavy workload.}
        \emph{Throughput and $99^{th}$ percentile latency results for RocksDB, Auto-tuned, SILK, and} \SYS.
    }
    \label{fig:usecase1-writeheavy}

    \vspace*{-5pt}
\end{figure*}

%% file: input/07-relatedwork.tex
\section{Related Work}
\label{sec:relatedwork}

Several SDS systems are targeted for specific I/O layers and storage contexts.
IOFlow~\cite{IOFlow:2013:Thereska}, sRoute~\cite{sRoute:2016:Stefanovici}, and PSLO~\cite{PSLO:2016:Li} tackle the virtualization layer. 
PriorityMeister~\cite{PriorityMeister:2014:Zhu} intercepts requests at the Network File System to enforce rate limiting services. 
At the block layer, Mesnier et al.~\cite{DifferentiatedServices:2011:Mesnier} classify requests and employ caching optimizations.
Pisces~\cite{Pisces:2012:Shue} and Libra~\cite{Libra:2014:Shue} enforce bandwidth guarantees under multi-tenant KVS. 
Malacology~\cite{Malacology:2017:Sevilla} improves the programmability of Ceph~\cite{Ceph:2006:Weil} to allow building custom applications on top of it.  
Retro~\cite{Retro:2015:Mace} and Cake~\cite{Cake:2012:Wang} implement resource management services at the Hadoop stack. 
\SYS advances these by being applicable over different I/O layers (we demonstrate this by implementing \SYS stages over RocksDB and TensorFlow), and providing a programmable and extensible library that allows developers to implement data plane stages with ease. 

SafeFS~\cite{SafeFS:2017:Pontes} and Crystal~\cite{Crystal:2017:Gracia} are the only two systems that share similar principles with \SYS, in terms of programmability and extensibility.
SafeFS provides a framework for stacking FUSE-based file systems on top of each other, each providing a different service to employ over requests.
Crystal extends the design of OpenStack Swift to implement custom services to be enforced over storage requests. 
However, both systems are bounded to a specific I/O layer (\emph{i.e.,} user-level file systems and OpenStack Swift), while \SYS ensures wider applicability. 
Moreover, as these systems do not allow additional layer information to be propagated to the data plane, they are unable to enforce policies at a finer granularity, such as those demonstrated in the RocksDB use case (\cref{subsec:tail-latency}).

%% file: input/08-conclusion.tex
\section{Conclusion}
\label{sec:conclusion}

In this paper we present \SYS, the first general-purpose SDS data plane framework.
It enables system designers to build custom-made data plane stages employable over different I/O layers.
\SYS provides differentiated treatment of requests and allows implementing fine-tuned storage services to cope with varied storage policies.

By combining ideas from SDS and context propagation, we demonstrate that \SYS allows decoupling system-specific I/O optimizations to a more programmable environment, promoting their portability and applicability to other systems and I/O layers, while also enforcing policies at a finer granularity.
We show this by implementing SILK's design principles in SDS fashion over RocksDB. 
Results show that a \SYS-enabled RocksDB improves tail latency at the $99^{th}$ by 4$\times$ under different workloads, and performs similarly to SILK.
Also, we demonstrate that by having global visibility of resources, \SYS-enabled deployments can achieve per-application dynamic bandwidth guarantees under a shared storage supercomputer environment.

%% file: input/09-acknowledgments.tex
\section*{Acknowledgments}

We thank the National Institute of Advanced Industrial Science and Technology for providing access to computational resources of AI Bridging Cloud Infrastructure (ABCI). 
We thank Vitor Enes and Cl\'{a}udia Brito for comments and suggestions. 
We also thank Oana Balmau for discussions about SILK.
This work was supported by the Portuguese Foundation for Science and Technology through a PhD Fellowship (SFRH/BD/146059/2019 -- Ricardo Macedo) and project PAStor (UTA-EXPL/CA/0075/2019 -- Jos\'{e} Pereira).
Work realized within the scope of the project BigHPC (POCI-01-0247-FEDER-045924 -- Jo\~{a}o Paulo), funded by the European Regional Development Fund, through the Operational Programme for Competitiveness and Internationalization (COMPETE 2020 Programme) and by National Funds through the Portuguese Foundation for Science and Technology, I.P. on the scope of the UT Austin Portugal Program.

%% file: input/10-availability.tex
\section*{Availability}

\SYS user-level library, along with tests and scripts used for conducting the experiments of this paper, are publicly available at \SYSurl.